\documentclass[amsmath,amssymb,aps,pre,letterpaper,superscriptaddress,floatfix,twocolumn,nofootinbib]{revtex4-2}
\usepackage{graphicx,xcolor} 
\usepackage{bm}
\usepackage{epstopdf}
\usepackage{textcomp}
\usepackage{gensymb}
\usepackage{array,multirow}
\newcolumntype{K}[1]{>{\centering\arraybackslash}m{#1}}
\usepackage{makecell}
\usepackage{blindtext}
\usepackage{verbatim}
\usepackage{xr}
\usepackage[normalem]{ulem}

\newcommand{\change}[1]{#1}

\renewcommand{\(}{\left(}
\renewcommand{\)}{\right)}

\usepackage{lipsum} % Required to insert dummy text

\graphicspath{{./figures/}}

\begin{document}

\title{Many Will Enter, Few Will Win: Cost and Sensitivity of Exploratory Dynamics
}

\author{Elena F. Koslover}
\affiliation{Department of Physics, University of California, San Diego, La Jolla, CA 92093, USA}
\author{Milo M. Lin}

\affiliation{Green Center for Systems Biology, University of Texas Southwestern Medical Center, Dallas, TX 75390}

\author{Rob Phillips}

\affiliation{Department of Physics and Division of Biology and Biological Engineering, California Institute of Technology, Pasadena, CA 91125}

\begin{abstract}
\noindent
A variety of biomolecular systems rely on exploratory dynamics to reach target locations or states within a cell.  Without a mechanism to remotely sense and move directly towards a target, the system must sample over many paths, often including resetting transitions back to the origin. 
We investigate how exploratory dynamics can confer an important functional benefit: the ability to respond to small changes in parameters with large shifts in the steady-state behavior. However, such enhanced sensitivity comes at a cost: resetting cycles require energy dissipation in order to push the system out of its equilibrium steady state. We focus on \change{minimalist models for} two concrete examples: translational proofreading in the ribosome and microtubule length control via dynamic instability to illustrate the trade-offs between energetic cost and sensitivity.
In the former, a \change{driven hydrolysis} step enhances the ability to distinguish between substrates and decoys with small binding energy differences. In the latter, resetting cycles enable catalytic control, with the steady-state length distribution modulated by sub-stoichiometric concentrations of a reusable catalyst.
 Synthesizing past models of these well-studied systems, we show how path-counting and circuit mapping approaches can be used to address fundamental questions such as the number of futile cycles inherent in translation and the steady-state length distribution of a dynamically unstable polymer. In both cases, a limited amount of thermodynamic driving is sufficient to yield a qualitative transition to a system with enhanced sensitivity, enabling accurate discrimination and catalytic control at a modest energetic cost.
\end{abstract}

\maketitle

\section*{Statement of Significance}
Living cells constantly burn energy even while maintaining a steady state. In this work, we explore some of the functional benefits of such energy consumption. We show how exploratory dynamics, in which molecular systems blindly sample many paths until they reach a target state, can convey qualitatively new abilities by enabling the system to respond to small differences in parameters. Such exploration requires dissipating energy to drive the resetting cycles. Using two example systems: proofreading in the ribosome and catalytic control of microtubule lengths, we investigate the trade-off between the energetic cost and the functional benefits of exploratory dynamics. Our results highlight how coupling an energy source to some transitions in a system can enhance its sensitivity to signals.

%\section*{Statement of Significance}
%
%Many biomolecular systems rely on exploratory dynamics, undergoing active resetting to the origin before reaching their target state. Such resetting cycles can enhance the sensitivity of the system, as explored here for two biologically relevant examples. Using a classic toy model for proofreading during protein synthesis, we highlight a simple probabilistic approach for computing ribosomal accuracy and energetic cost. Even in the presence of excess decoy tRNAs, a very small number of futile reset cycles can allow the system to approach close to the minimal error rate. For dynamically unstable microtubules, energy input allows the steady-state length to be sensitive to small concentrations of a passive enzymatic co-factor, enabling catalytic control of structure at the cellular scale. 

\section*{Overview}

\noindent It is our great pleasure to contribute to this special issue dedicated to the life and work of Prof. Erich Sackmann. 
Sackmann was a pioneer in the broad field dedicated to using
the tools of physics to understand the fascinating phenomena of life.
One of the most intriguing aspects of living organisms that makes them so different from their inanimate counterparts is the expenditure of energy to maintain nonequilibrium steady states. Many of the phenomena that exploit such energy consumption share features of exploratory dynamics: sampling many paths with occasional resetting en route to a target state.
Common examples include biological polymerization in processes such as replication, translation, and cytoskeletal filament dynamics. In this paper, we examine two case studies (protein translation and microtubule growth), linking the energetic cost \change{to functional benefits in terms of accuracy and sensitivity}.

\section{Introduction}
\noindent

One of the fundamental challenges faced by living cells is the need to carry out their functions while `blind' -- lacking a centralized omniscient organizer that can direct cellular components where to go and which interactions to perform in what order. \change{While some cellular processes can be guided via signal gradients~\cite{wollman2005efficient,fuller2010self} or prearranged cytoskeletal transport highways~\cite{burute2019cellular,agrawal2022morphology}, many rely on exploring a broad set of pathways rather than heading directly to the target. Whether searching through physical space or through chemical state-space, exploratory dynamics~\cite{gerhart1997exploratory_ch4} follows a characteristic pattern: repeatedly sampling through the available states, fixing and amplifying the targets when a trajectory stumbles across them, and resetting unsuccessful trajectories to try again.}

Such exploratory behavior was first highlighted in the context of dynamically unstable microtubules that engage in search-and-capture cycles of random growth and rapid depolymerization to find chromosomes when constructing the mitotic spindle~\cite{KirschnerSearch1986,Heald2015}. The same principle applies to other spatial search problems in the cell. For example, the ability of neurons to localize mitochondria in regions of high energy demand~\cite{misgeld2017mitostasis} relies on `sushi-belt' dynamics~\cite{williams2016dendritic}, where a motile population of organelles constantly cycles via motor-driven transport through neuronal projections, with regions of enhanced stopping encoded through local increases in calcium~\cite{macaskill2009miro1} or glucose concentrations~\cite{pekkurnaz2014glucose}. In the endocytic pathway, endosomes bearing activated receptors must wander through the cellular periphery until they encounter other organelles that trigger phosphoinositide conversion and deactivate the signal~\cite{york2023deterministic,villasenor2016signal}.

Analogous exploratory processes occur on the molecular scale, particularly in the context of sensing and proofreading. First introduced to explain the accuracy of polymerization-based information transfer~\cite{hopfield1974kinetic,ninio1975kinetic}, kinetic proofreading relies on chaining together intermediate states that the system must pass through before reaching a target. By providing multiple opportunities to reset back to the origin from each intermediate state, such proofreading can amplify the likelihood of following a pathway involving `right' versus `wrong' interactions. Thus, a ribosome searches for the next amino acid to add to a growing polypeptide chain by exploring through  intermediate states that might involve the right or wrong tRNA. The higher rate of resetting (release of the tRNA) for the wrong amino acid allows a greater probability that the final step of peptide elongation is reached only with the correct amino acid. 
Similar exploration through intermediate states shows up in sensory systems, such as T-cell activation~\cite{cui2018identifying} or chemotactic signaling~\cite{hathcock2024time}. In both cases, incorrect ligands that bind weakly to receptors are more likely to be released, resulting in resetting during each intermediate step.
Proofreading through resetting is also thought to contribute to the self-assembly of large multimeric structures, including sequence-specific RecA filament formation on DNA~\cite{bar2002protein} and viral RNA packaging~\cite{mizrahi2022spanning}. 

Protein quality control systems provide additional examples of exploratory dynamics that leverage resetting to accurately sort components into distinct pathways.
The ubiquitinating enzyme APC is able to distinguish its substrates among myriads of decoy proteins by sequentially marking multiple lysine groups on the target protein, eventually triggering the degradation of the substrate~\cite{lu2015specificity}. The resulting difference in APC binding affinity on ubiquitinated substrates versus non-ubiquitinated decoys makes it more likely that only the correct proteins are targeted for degradation.
 In the secretory pathway, newly translated proteins are tagged by the addition of glycan chains that facilitate binding to chaperones which help fold the proteins~\cite{adams2019protein}. Multiple cycles of glycosylation  in the ER allow chaperones to make several attempts at folding a nascent protein before it proceeds towards the terminal pathways of export or degradation~\cite{brown2021design}. 

 Given the prevalence of resetting dynamics in intracellular systems, a natural question is why this approach is so common and what advantages it might offer to the cell. %Broadly speaking, cells contend against constraints in space, time, and information. When the location of the target state is not known a priori, it is not possible to construct a landscape that funnels the system directly to the target. If the system has to explore over a relatively flat landscape with many accessible states, resetting to the origin can greatly speed up the search time~\cite{evans2011diffusion,evans2020stochastic}. 
 The effects of resetting on speed in subcellular exploration are considered in previous work ~\cite{evans2011diffusion,evans2020stochastic} and addressed in a cohesive framework in another article within this issue~\cite{Kondev2025}. 

 Instead of considering the temporal advantages, here we illustrate how exploratory dynamics enhances sensitivity: the ability of the cell to respond to small differences in system parameters. We contend that seemingly distinct functional objectives, including concerted activation of specific signaling molecules, accurate discrimination of targets from decoys, and regulation of molecular assembly size, are all manifestations of the same phenomenon: namely that exploratory dynamics magnifies the effect of changing system parameters (inputs) on steady-state observables (outputs). Such sensitivity comes at a cost of energy expenditure because resetting necessarily involves cycles that break detailed balance. 
Using kinetic proofreading and microtubule length regulation as examples, we show how signal gain depends on energy dissipation and the concomitant driving of systems away from equilibrium detailed balance toward non-equilibrium exploratory dynamics. Importantly, we show that energy expenditure through driving one transition can qualitatively change the response function to parameters elsewhere in the system.

 One manifestation of this phenomenon is catalytic control -- a ubiquitous feature of biomolecular regulatory systems, where a reusable catalyst (such as a kinase) alters the state of its substrates. 
Because a catalyst modifies transition barriers only, the input parameter (catalyst concentration) can have no effect on steady-state output under equilibrium conditions. 
 However, in the presence of driven exploratory dynamics, the steady-state probability of different substrate states can in fact become dependent on the level of catalyst present.
 This feature enables a small (sub-stoichiometric) number of regulatory molecules to trigger large-scale changes in cellular state. Examples include the whole-sale ubiquitination and degradation of specific classes of proteins during mitosis~\cite{lu2015specificity} and the catalytic regulation of microtubule length during cell division~\cite{belmont1990real}. Such catalytic control requires energy dissipation somewhere in the system, even when the catalysis step itself does not couple to an external energy source.

In this paper, we investigate how exploratory dynamics with resetting allows biochemical systems to enhance their sensitivity. We consider both input sensitivity against decoy signals, as well as output sensitivity to sub-stoichiometric catalytic regulation.
For concreteness, we focus on two key biological examples that illustrate these features: single-step kinetic proofreading in ribosomal translation, and multi-step resetting in microtubules undergoing dynamic instability. Along the way, we highlight two pedagogically useful approaches to describing such systems: path-counting (which intuitively incorporates exploratory dynamics), and circuit mapping (which clarifies the relation between energetic driving and steady-state distributions). These systems highlight both the functional benefits of resetting dynamics and the concomitant cost in energy dissipation.

\section{Kinetic Proofreading in the ribosome}

The notion of accuracy can be defined for stochastic reaction systems where there are multiple terminal states of which only a particular subset is considered `correct'. 
For systems at equilibrium, the error rate (ratio of wrong to right pathways selected) is bounded by the difference in free energy change from the initial to the final state~\cite{sartori2015thermodynamics}. 
\change{
In situations where the binding energy for wrong versus right substrates is similar, and where there is an excess concentration of the decoy substrate, purely thermodynamic discrimination may result in unacceptably high error rates. For example, the difference in binding energy for cognate or near-cognate tRNA during ribosomal translation is expected to be on the order of 1-2 hydrogen bonds, or only a few $k_bT$~\cite{ogle2002selection}. Given the concentration excess due to multiple possible near-cognate substrates, the thermodynamic error rate would be expected to be above $5^\%$. On the scale of a 300-amino-acid long peptide, being able to build an error-free chain even half of the time would require error rates below $0.2\%$, necessitating the introduction of an alternate proofreading mechanism.

Many biological systems increase their accuracy by introducing multiple intermediate states, with the probability of resetting to the origin higher for pathways leading to the wrong terminal state --
 a process that has been termed `kinetic proofreading'~\cite{hopfield1974kinetic,mellenius2017transcriptional,banerjee2017elucidating,murugan2012speed,cui2018identifying}. 
This approach to proofreading via exploratory dynamics can also include discrimination in the forward rates for moving to the next intermediate~\cite{banerjee2017elucidating}. %as well as in the release rates for resetting.
But even for the case where discrimination is localized entirely to the release steps, systems with actively driven resetting cycles can combine together the distinct binding energies of multiple intermediate states, surpassing purely thermodynamic limits on accuracy~\cite{murugan2012speed}.}

A number of theoretical works have sought to elucidate the connections between energy dissipation, speed, and accuracy of a proofreading system~\change{\cite{Lan2012,murugan2012speed,banerjee2017elucidating,yu2022energy}}. Even for a simple single-step enzyme, accurately distinguishing between cognate and non-cognate substrates requires product formation to be slow compared to the binding-unbinding equilibration~\cite{tawfik2014accuracy}. From the same principle, the forward step in multi-stage proof-reading pathways must be arbitrarily slow to reach the minimal possible error rate, highlighting a trade-off between speed and accuracy~\cite{johansson2008rate}. Furthermore, when proofreading relies on resetting, each such resetting cycle implies a slow-down in the total time to reach the target. For multi-step pathways, the time to reach a target scales exponentially with the number of states if resetting is more likely than forward stepping, and linearly otherwise~\cite{murugan2012speed}. Since each intermediate state provides an extra opportunity for proofreading, accuracy comes at a cost in speed. \change{In ribosomal translation, measured kinetic parameters have been shown to place the system in a regime of near-optimal speed at the expense of about ten-fold higher error rate~\cite{banerjee2017elucidating}. } The trade-off between speed and accuracy has also been noted in models of T-cell activation~\cite{cui2018identifying} and chemosensory receptor arrays~\cite{Lan2012,hathcock2024time}.

\change{A proofreading system that relies on exploratory dynamics must balance not only the speed and accuracy but also the energetic cost of resetting cycles. Notably, some kinetic parameters for ribosomal translation appear be tuned so as to decrease the overall energy cost while allowing a minor increase in speed~\cite{banerjee2017elucidating}, highlighting the importance of considering energetic constraints on proofreading.}
The energetic cost depends on both the dissipation per turn of the resetting cycle and the typical number of such cycles before  reaching the target state.
 Thus, for translational proofreading we could ask how many tRNAs are released from the \change{high-energy intermediate} state before an amino acid is successfully incorporated into the growing polypeptide chain. In other words, given the presence of excess wrong versus right tRNAs, how many GTP must be hydrolyzed per elongation event?
This question has been formulated in terms of the total entropy production (dissipation rate normalized by the incorporation rate)~\cite{sartori2015thermodynamics}, and in terms of the number of futile cycles~\cite{banerjee2017elucidating, yu2022energy}. Some fundamental bounds have been proposed for relating the dissipation and the error rate of a proofreading system. At equilibrium, the error rate is equal to the ratio of Boltzmann factors for the right versus wrong products~\cite{phillips2012physical}. For non-equilibrium proofreading systems, the error rate can be driven up or down by an exponential factor incorporating both the total entropy production and the `excess work' put into the system beyond the overall free energy difference for incorporation~\cite{sartori2015thermodynamics}. For multistage proofreading schemes, the minimal energy cost necessary to sustain a particular error rate decreases with both the number of intermediate states and the right-vs-wrong discrimination factor for resetting from each intermediate state~\cite{yu2022energy}.

\change{
Below we seek to provide a pedagogically helpful analysis of the energy cost versus accuracy trade-off for a classic single-intermediate model of translational proofreading. Our emphasis here is not on faithfully reproducing the detailed kinetics of ribosomal translation, but rather on demonstrating the principle that driven exploratory dynamics can increase the sensitivity of a proofreading system.
We begin with a model incorporating irreversible transitions, using a splitting-probability approach to compute the number of excess GTP hydrolyzed per elongation event. More complex reaction schemes, \change{including discrimination in all rate constants}, can also be reduced to this simple model. Notably, the ability to compute energy costs for a system with irreversible transitions is important because reverse processes are often so rare as to be never observed. Experimentally parameterized kinetic models thus often contain irreversible arrows~\cite{rodnina2001ribosome,wohlgemuth2011evolutionary,zaher2010hyperaccurate}.}

We then expand our approach to incorporate fully reversible steps and demonstrate how increasing energetic driving causes the system to transition between distinct regimes, with intermediate driving strength but large numbers of futile cycles required to achieve the greatest accuracy. Overall, active driving enhances the sensitivity of translational proofreading to small changes in tRNA binding energy, enabling it to accurately discriminate between correct and wrong amino acids for incorporation into the peptide chain.

\subsection{Irreversible Model}

\begin{figure}
\includegraphics[width=8.3cm]{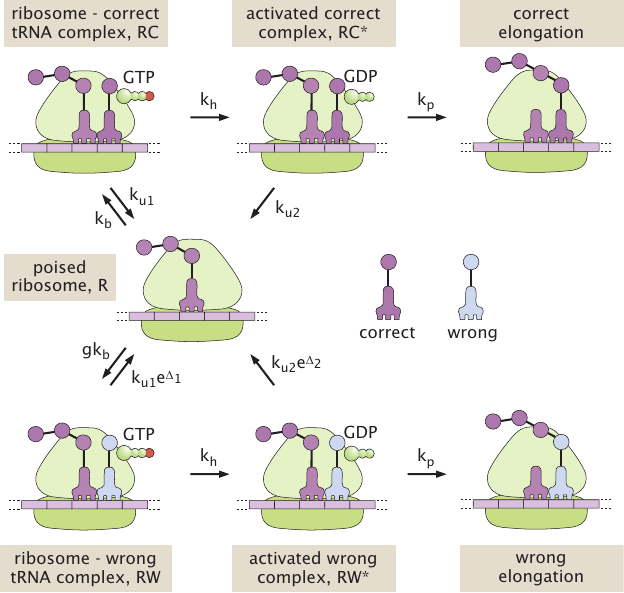}
\caption{\textsf{\textbf{Schematic of a classic 2-step kinetic proofreading model for translation.}} The states are: $R$ = empty ribosome, $RC$ = ribosome with correct tRNA bound, $RW$ =  ribosome with wrong tRNA bound, $RC^*, RW^*$ = ribosome with correct or wrong tRNA bound in high-energy state after GTP hydrolysis.   } 
\label{fig:scheme}
\end{figure}

As shown in Fig.~\ref{fig:scheme}, we represent translational elongation via a classic 
simplified reaction scheme, as proposed by Hopfield in his seminal work on kinetic proofreading~\cite{hopfield1974kinetic}. This scheme begins with an empty ribosomal binding site (denoted as the R state). A reversible binding step allows a tRNA loaded with the correct (C) or wrong (W) amino acid to interact with the template strand. 
  For simplicity, we assume that the tRNA arrival is diffusion-limited, so that the binding rate for the correct amino acid is $k_b$ and the rate for the wrong amino acid is $gk_b$. 
  \change{The factor $g$ accounts for the excess concentration of wrong versus correct tRNAs. There are up to $61$ possible tRNAs, with 1-4 of them carrying any given amino acid~\cite{kramer2007frequency}. These tRNAs are present in varying concentrations~\cite{dong1996co} in the cell, and will have varying degrees of interaction with a particular codon on the template strand.
  For the purposes of this model, we are interested only those tRNAs capable of interacting with the ribosome sufficiently to proceed through subsequent steps in the cycle. Experimental observations indicate that only near-cognate tRNAs have a measurable chance of doing so~\cite{kramer2007frequency,joshi2019problem}, and we estimate a typical concentration excess of $g=4$ for near-cognate tRNAs carrying the wrong amino acid (see Appendix A).}

  Even in the absence of proofreading, we would expect some discrimination between tRNAs carrying right and wrong amino acids based on their different binding energies. The thermodynamic limit on accuracy~\cite{sartori2015thermodynamics} is then set by the equilibrium ratio of wrong versus correct tRNAs bound to the ribosome, expressed as $f_\text{passive} = ge^{-\beta \Delta \varepsilon}$, where $\Delta \varepsilon=\varepsilon_w -\varepsilon_c$ is the difference in binding energies,
 and we adopt the notation $\beta=1/k_BT$. 
For much of what follows, we will use the convention of defining dimensionless energies such as $\Delta_1=\beta \Delta \varepsilon$.
Given these conventions, we set the off-rates for the reversible interaction to be $k_{u1}$ and $k_{u1}e^{\Delta_1}$ for the correct and wrong amino acids, respectively.

\begin{figure*}
\includegraphics[width=\textwidth]{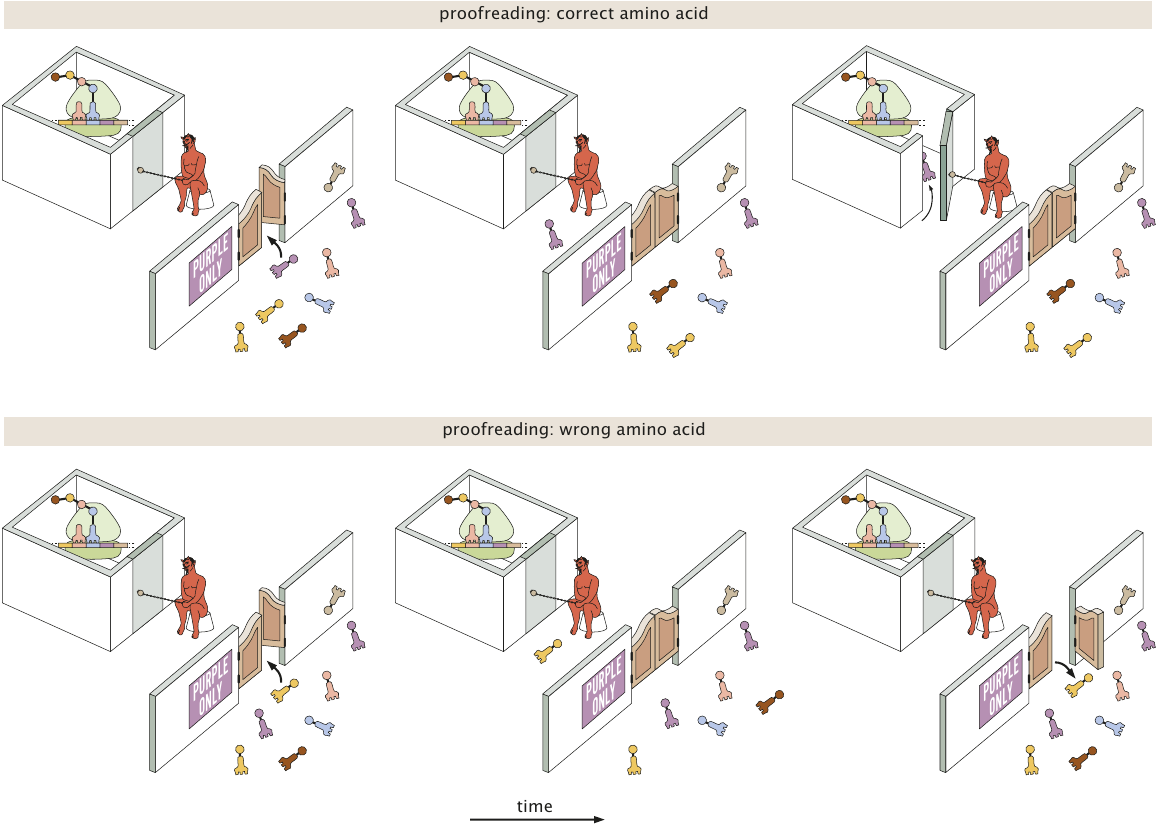}
\caption{\textsf{\textbf{Pictorial description of the two-step process whereby correct and wrong tRNA-amino acid pairs are distinguished.}}  The first step is passive and includes a preference for one type of tRNA-amino acid pair over all the others.  The second energetically costly active step provides a second chance to distinguish correct and wrong tRNA-amino acid pairs. }
\label{fig:cartoon}
\end{figure*}

In a somewhat whimsical analogy (illustrated in Fig.~\ref{fig:cartoon}), we can describe translational elongation as entry of visitors to a particularly selective clubhouse. The first reversible binding step might then correspond to a swinging door with a passive sign declaring who can come in. This passive filter allows for some discrimination, but does not completely keep out unwelcome visitors who might sneak through the swinging door. As in protein translation, a subsequent active step to check the visitors' identity is needed.

The next step in the translational elongation pathway involves the hydrolysis of GTP (in the EF-Tu cofactor associated with the tRNA). This hydrolysis serves as a tightly-coupled energy source, transitioning the ribosome to a high-energy intermediate state (RC$^*$ or RW$^*$). We assume the same hydrolysis rate $k_h$ regardless of which amino acid is present (\change{see Appendix A for a model without this assumption}). Because of the large free energy change associated with GTP hydrolysis and phosphate release, this step is taken to be effectively irreversible.

The \change{hydrolyzed intermediate} provides an opportunity for proofreading in that the tRNA can again disassociate from the ribosome, essentially serving as a reset in the overall exploratory dynamics. The release process is discriminatory, with the correct tRNA falling off at rate $k_{u2}$ and the wrong one at rate $k_{u2} e^{\Delta_2}$. The quantity $\Delta_2$ can correspond to either the difference in binding energy between the right and wrong tRNAs, or to a difference in the barrier heights for dissociation~\cite{sartori2013kinetic}. Either way, we assume this \change{intermediate} state is so high on the energy landscape that the dissociation process is irreversible. For the dissociated tRNA to return to the ribosome, it must again pass through the reversibly bound state.

In our analogy, the proofreading step corresponds to an active identity check at the inner door of the clubhouse (Fig.~\ref{fig:cartoon}, rightmost panels). Such a step is costly in that it requires an energy-consuming ``Maxwell's Demon" to open the inner door selectively for the correct visitors. However, it has the advantage of more accurately vetting which visitors are allowed to enter the clubhouse. The sequential passive then active filtering steps, allow for high accuracy to be achieved without overworking the demon, since the number of undesired visitors attempting to sneak through the swinging door is already reasonably low.

Finally, there is an irreversible step for forming the peptide bond to incorporate the new amino acid into the growing peptide chain. We assume this step occurs with rate $k_p$, regardless of the amino acid identity. In our analogy, this corresponds to the final step of visitors being permanently sworn into the exclusive club.

\change{It is important to note that peptide bonds themselves are thermodynamically unstable, so that long peptide lifetimes depend on these bonds being kinetically trapped~\cite{england2013statistical}. In order for the final peptide formation step to be  irreversibly trapped, it needs to start from a high-energy intermediate ($RC^*, RW^*$ states). The GTP hydrolysis step is then necessary to allow the system to reach this high energy state. Thus the energy input into translation plays dual roles in both biosynthesis of unstable bonds and enhancing accuracy through proofreading.}

\change{We begin our analysis with the original minimalist Hopfield model for pedagogical purposes, as it provides one of the simplest possible systems for proofreading via resetting. In reality, ribosomal translation has been shown to discriminate between cognate and non-cognate tRNAs in the rate constants for hydrolysis and elongation, as well as in unbinding and release~\cite{rodnina2001ribosome,wohlgemuth2011evolutionary,zaher2010hyperaccurate}. Furthermore, translation includes additional reversible and irreversible steps, such as codon recognition and GTPase activation. In Appendix~A, we show how the analysis can be expanded to accommodate more complex single-stage proofreading schemes, as well as deriving the corresponding effective parameters $\Delta_1, \Delta_2$, which represent the discrimination factors inherent in the system. As described in the Appendix, the relevant parameter estimates for ribosomal translation are: $g \approx 4, \Delta_1 \approx 4, \Delta_2 \approx 8$.}

\subsection{Model solution}

The kinetic scheme illustrated in Fig.~\ref{fig:scheme} represents a network of Markovian transitions between states. \change{The ribosome can be thought of as hopping stochastically on this network, with a matrix of splitting probabilities determining the probability of each outward hop out of any given state. The system properties can be analyzed by directly manipulating this splitting probability matrix~\cite{banerjee2017elucidating}. However, for purposes of conceptual clarity, we take an alternate `blackboard-friendly' approach which computes the probabilistic weight of each possible path describing a ribosome-tRNA interaction.}

  We are interested in
the average cost of translation in the presence of excess ``wrong" tRNA/amino acid pairs.
More precisely, starting in the empty state (R), how many hydrolysis events must occur before the system reaches a terminal state, elongating the peptide chain by an additional amino acid? This provides a measure of the energetic cost per amino acid for building a nascent peptide chain. The energetic cost can then be compared to the fidelity of translation, expressed as the error rate $f$ (ratio of wrong versus correct amino acids incorporated into the chain).

As the system hops between the discrete states, there is a choice at each step of which outward arrow to follow.
The splitting probabilities at each state can be obtained from the ratio of rates on the outward arrows.
Starting from the poised R state, the probabilities of binding the correct tRNA ($p_{bc}$) or the wrong tRNA ($p_{bw}$)  are given by
\begin{equation}
	\begin{split}
		p_{bc} = \frac{1}{1+g}, \quad 	p_{bw} = \frac{g}{1+g}.
	\end{split}
	\label{eq:pbcw}
\end{equation}
The probabilities of unbinding ($p_{u1c}, p_{u1w}$) or  undergoing hydrolysis ($p_{hc}, p_{hw}$) from the RC and RW states, respectively, are given in turn by
\begin{subequations}
    \begin{align}
		p_{u1c} = \frac{k_{u1}}{k_{u1}+k_h}, & \quad p_{u1w} = \frac{k_{u1}e^{\Delta_1}}{k_{u1}e^{\Delta_1}+k_h},	\\
		p_{hc} = \frac{k_h}{k_{u1}+k_h}, & \quad 	p_{hw} = \frac{k_h}{k_{u1}e^{\Delta_1}+k_h}.
    \end{align}
\end{subequations}

Similarly, from the \change{ hydrolyzed intermediate} states, the probabilities of unbinding ($p_{u2c},p_{u2w}$) or peptide elongation ($p_{pc},p_{pw}$) are given by
\begin{subequations}
    \begin{align}
		p_{u2c} = \frac{k_{u2}}{k_{u2}+k_p}, & \quad p_{u2w} = \frac{k_{u2}}{k_{u2}e^{\Delta_2}+k_p},	 \\		 
        p_{pc} = \frac{k_p}{k_{u2}+k_p}, & \quad 	p_{pc} = \frac{k_p}{k_{u2}e^{\Delta_2}+k_p}.
	\end{align}
	\label{eq:pu2}
\end{subequations}
We note that these are splitting probabilities for a ribosome assumed to be starting in a specific state; they do not directly include the steady-state probability of being in that state or give the resulting flux along an arrow.
Because the system is Markovian, we can define a probabilistic weight for any multi-step path (from a given starting state) by multiplying the probabilities of all the steps. The weights of different paths can then be added together. 

\change{There are many possible paths that lead to a final state, including any number of unbinding or release cycles before finally proceeding to hydrolysis or elongation. Accounting for all these paths is made easier if we break them up into individual ``interaction events", consisting of simple loops that start and end in the empty ribosome state R, with no intervening visits to that state. Each interaction event must proceed through one of only 6 possible paths: the system returns to the original state through either unbinding, release, or elongation, with either a correct or wrong tRNA in each case. Every path to elongation is composed of one or more such interaction loops and every interaction is independent of the proceeding one. Hence, the overall behavior of the system can be computed based on the outcome probabilities for each individual interaction.
}		
	
The probability that an interaction includes a hydrolysis event is the sum of two terms, corresponding to paths with correct or wrong binding and results in the expression
\begin{equation}
	\begin{split}
		p_\text{hyd} = p_{bc} p_{hc} + p_{bw}p_{hw}.
	\end{split}
\end{equation}
Similarly, the probability of a ribosome-tRNA interaction resulting in elongation is given by
\begin{equation}
	\begin{split}
		p_\text{el} = p_{bc} p_{hc} p_{pc}+ p_{bw}p_{hw}p_{pw}.
	\end{split}
\end{equation}

We want to calculate the average number of hydrolysis events preceding an elongation step. 
This can be done quite simply using conditional probabilities, a unifying concept that has been recently highlighted for its utility in clarifying the exploratory dynamics of biological processes~\cite{hachmo2023conditional}.
The  problem at hand is equivalent to a classic probability question: ``If you throw a fair 6-sided die until you first get a 6, how many throws of the dice do you need on average if you only count the throws with even faces?". Since only even throws are counted, we need the probability of success (rolling a 6), conditional on the roll being even. This conditional probability is given by $p = P(6|\text{even}) = 1/3$. The average number of even rolls to reach a 6 (including the last one) is then $1/p = 3$.

For the proofreading case, the conditional probability of a ribosome-tRNA interaction leading to elongation, given that a hydrolysis event occurs, is $p = P(\text{elongation} | \text{hydrolysis}) = p_\text{el} / p_\text{hyd}$. We therefore compute the average number of futile hydrolysis events (not counting the one that successfully leads to elongation) as
\begin{equation}
	\begin{split}
		\left<n\right> & = \frac{1}{p}-1 = \frac{p_{bc} p_{hc} (1-p_{pc})+ p_{bw}p_{hw}(1-p_{pw})}{p_{bc} p_{hc} p_{pc}+ p_{bw}p_{hw}p_{pw}} \\
		& = \frac{p_{bc} p_{hc} p_{u2c}+ p_{bw}p_{hw}p_{u2w}}{p_{bc} p_{hc} p_{pc}+ p_{bw}p_{hw}p_{pw}}.	
	\end{split}
	\label{eq:avgn}
\end{equation}
Notably, this average number of futile cycles can also be intuitively expressed as the ratio of unsuccessful to successful hydrolysis events: $\left<n\right> = (1-p)/p = P(\text{not elongation}|\text{hydrolysis}) / P(\text{elongation}|\text{hydrolysis})$.
Plugging in Eq.~\ref{eq:pbcw}-~\ref{eq:pu2} gives the result in terms of kinetic parameters:
\begin{widetext}
	\begin{equation}
		\begin{split}	
			\left<n\right> & 
			= \frac{k_{u2}}{k_p} \left[\frac{\(k_{u1}e^{\Delta_1} + k_h\)\(k_{u2}e^{\Delta_2} + k_p\) + ge^{\Delta_2}\(k_{u1} + k_h\)\(k_{u2} + k_p\)}{\(k_{u1}e^{\Delta_1} + k_h\)\(k_{u2}e^{\Delta_2} + k_p\) + g\(k_{u1} + k_h\)\(k_{u2} + k_p\)}\right]
		\end{split}
		\label{eq:excesshyd}
	\end{equation}
\end{widetext}

The error rate can similarly be expressed  as the ratio of probabilistic weights for \change{interactions} with the wrong tRNA leading to elongation versus \change{interactions} with the correct tRNA doing so:
\begin{equation}
	\begin{split}
		f & = \frac{p_{bw} p_{hw} p_{pw}}{p_{bc} p_{hc} p_{pc}}	 = g\(\frac{k_{u1}+k_h}{k_{u1}e^{\Delta_1}+k_h}\)\(\frac{k_{u2}+k_p}{k_{u2}e^{\Delta_2}+k_p}\).
	\end{split}
	\label{eq:fidelity}
\end{equation}

 \change{The expressions for $\left<n\right>$ and $f$ in terms of splitting probabilities can also be applied to more general kinetic models. In Appendix A, we describe a general scheme 
 	that includes both the Hopfield model in Fig.~\ref{fig:scheme} and  more detailed models that incorporate codon recognition, GTPase activation, etc, while allowing for discrimination between wrong and correct tRNAs in all transition rates~\cite{rodnina2001ribosome, wohlgemuth2011evolutionary}.}

\subsection{Accuracy and cost in irreversible model}
\noindent

The energetic cost of translational elongation can be computed from the number of hydrolysis events, as $\Delta E = (\left<n\right>+1)\epsilon_\text{GTP}$ where $\epsilon_\text{GTP}$ is the change in free energy associated with hydrolyzing one ATP molecule. We note some features of this energy cost are apparent from Eq.~\ref{eq:excesshyd}. First of all, the energy cost is not dependent on the binding rate $k_b$ of the tRNAs, as expected since binding initiates all interactions, whether they are futile or not.

If the release step is indiscriminate between right and wrong amino acids ($\Delta_2 \rightarrow 0$), then the number of excess hydrolysis events becomes $k_{u2}/k_p$ (the rate of release from the \change{hydrolyzed} state, relative to the rate of peptide bond formation). 
\change{This is expected since the ratio of futile versus productive exits from the hydrolyzed states is given by their ratio of rates, and this should equal the average number of futile exits $\left<n\right>$ per each productive one.}
In this limit, the energetic cost of elongation can be minimized by preventing release from the \change{hydrolyzed} state ($k_{u2} \rightarrow 0$) or effectively removing the indiscriminate proofreading step from the system. The same limit is obtained when there are no wrong amino acids present in the system ($g\rightarrow 0$). 

\begin{figure}[t]
	\centerline{\includegraphics[width=8.3cm]{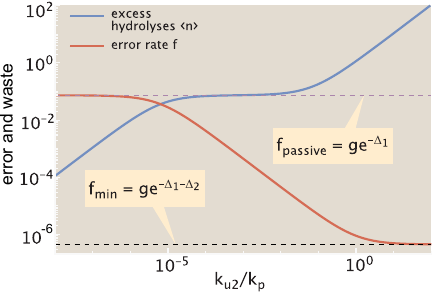}}
	\caption{\textsf{\textbf{Proofreading error is decreased at the cost of additional futile cycles}} Error rate (red, $f$) decreases and the cost in excess hydrolysis events (blue, $\left<n\right>$) increases as the ratio of release to elongation rates ($k_{u2}/k_p$) is raised. Dotted purple line marks the passive error rate in the absence of proofreading. This is also equal to the excess hydrolysis count in the  intermediate regime where wrong tRNAs are likely to dissociate during the proofreading step while correct tRNAs are likely to proceed to elongation.  Dotted black line marks the fundamental fidelity limit for this proofreading system, which can only be reached with an infinite number of hydrolyses per elongation.
		Results shown are in the limit $k_h \rightarrow 0$, with $\Delta_1 = 4, \change{\Delta_2 = 8, g = 4}$. }
	\label{fig:avgn}
\end{figure}

We next consider the limit where hydrolysis is much slower than the unbinding rate ($k_h \ll k_{u1}$), allowing the initial binding step to equilibrate before each hydrolysis occurs. This limit was also assumed in past analyses of kinetic proofreading processes~\cite{hopfield1974kinetic,murugan2012speed}. In this case, the average count of excess hydrolyses, and the error rate simplify to
\begin{subequations}
	\begin{align}
\left<n\right> & \xrightarrow{k_h\rightarrow 0} \frac{k_{u2}}{k_p} \left[ \frac{e^{\Delta_1}\left(k_{u2} e^{\Delta_2} + k_p\right) + g e^{\Delta_2} \left(k_{u2} + k_p\right)}
{ e^{\Delta_1}\left(k_{u2} e^{\Delta_2} + k_p\right) + g \left(k_{u2} + k_p\right)}
\right], \\	
f & \xrightarrow{k_h\rightarrow 0} g e^{-\Delta_1}\left(\frac{k_p+k_{u2}}{k_p+k_{u2}e^{\Delta_2}}\right). \label{eq:f_smallh}
	\end{align}
\end{subequations}
In this limit, neither the energy cost nor the fidelity are dependent on the unbinding rate $k_{u1}$. Instead, they are determined by the discrimination energy for the initial binding step ($\Delta_1$), and for dissociation during the proofreading step ($\Delta_2$), as well as the fold-excess of wrong amino acids ($g$) and the relative rate of release during proofreading ($k_{u2}/k_p$). Both error rate and the excess hydrolysis count are plotted in Fig.~\ref{fig:avgn}, as a function of this release rate.

In the regime where the release rate during proofreading is very low ($k_{u2}\ll k_p$), there are very few excess hydrolysis events ($\left<n\right>\rightarrow 0$). However, the system also loses its ability to proofread since no incorrect tRNAs are released after they enter the \change{hydrolyzed} state. The error rate then approaches the thermodynamic limit for passive binding: $f_\text{passive} = g e^{-\Delta_1}$. 

As the release rate increases, we enter an intermediate regime where the correct tRNA has a low chance of being released during proofreading but the wrong tRNA has a high chance of being released: $k_{u2}\ll k_p \ll k_{u2} e^{\Delta_2}$. In addition, we also assume $g e^{-\Delta_1} \ll \frac{k_{u2}}{k_p} e^{\Delta_2}$, implying that the probability of \change{post-hydrolysis} release for wrong amino acids is high enough to overcome the error inherent in the initial binding.
Since most wrong tRNAs are then released after activation, the average number of excess hydrolysis events is equal to the (thermodynamic) error rate of the hydrolysis process itself: $\left<n\right> \approx g e^{-\Delta_1}$. Notably, this quantity is independent of $k_{u2}$, giving rise to a plateau region where the excess hydrolysis count remains flat while the error rate continues to decrease (Fig.~\ref{fig:avgn}).

In the regime where the release rate is high for all tRNAs ($k_{u2}\gg k_p$), the error rate approaches its fundamental limit~\cite{hopfield1974kinetic} of $f_\text{min} =  ge^{-\Delta_1 - \Delta_2}$. However, in this regime the number of excess hydrolysis cycles approaches infinity as many release events precede each successful incorporation.

\begin{figure}[t]
	\centerline{\includegraphics[width=8.3cm]{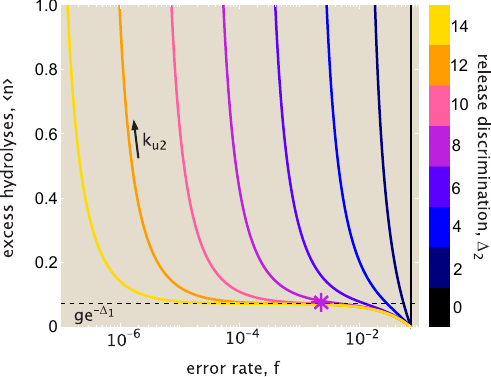}}	
	\caption{\textsf{\textbf{Limited energetic cost allows for near-optimal fidelity.}} The average number of excess hydrolysis events $\left<n\right>$ is plotted versus the error rate $f$. Each curve corresponds to a fixed value of $\Delta_2$, with increasing release rates $k_{u2}$ moving to the left along the curve. Dashed black line marks the thermodynamic error rate for the hydrolysis itself, corresponding to the probability that the wrong tRNA reaches the \change{hydrolyzed intermediate} state. Large values of the release discrimination energy $\Delta_2$ allow for the plateau value of excess hydrolysis to extend to low overall error rates.
		Results shown are in the limit $k_h \rightarrow 0$, with $\Delta_1 = 4, \change{g = 4}$. \change{Asterisk marks cost and error rate values for realistic parameters, as derived in Appendix A.} }
	\label{fig:avgnf}
\end{figure}

A direct relation between the error rate and the number of excess hydrolysis events, in the limit of $k_h\rightarrow 0$, can be written as
\begin{equation}
\begin{split}
\left<n\right> = \(\frac{f_\text{passive} - f}{f-f_\text{min}}\)\(\frac{f_\text{min} + f f_\text{passive}}{f_\text{passive}+f f_\text{passive}}\),
\end{split}
\end{equation} 
where the error rate is always constrained to lie in the range $f_\text{min} < f < f_\text{passive}$. For the case where $\Delta_1 = \Delta_2$, this expression is identical to that previously derived for the minimum number of futile hydrolysis cycles in a reversible proofreading system with a fixed error rate~\cite{yu2022energy}.

From this relation (plotted in Fig.~\ref{fig:avgnf}), we again see the three regimes for the error rate. For the lowest error rates ($f \rightarrow f_\text{min}$), the proofreading process becomes exceedingly wasteful as $\left<n\right>$ goes to infinity. For the highest error rates ($f\rightarrow f_\text{passive}$), the excess hydrolysis count goes to 0: there is no waste, but the system also loses the accuracy boost due to proofreading. In the intermediate plateau regime, corresponding to
$f_\text{min} \ll \left\{ff_\text{passive}, f\right\} \ll f_\text{passive}$,  
 the hydrolysis count is approximated by $\left<n\right> \approx g e^{-\Delta_1}$. The plateau becomes wider, accessing lower error rates, when there is greater discrimination for release of correct versus wrong tRNAs (higher $\Delta_2$, or larger ratio of $f_\text{passive}/f_\text{min}$).

The plateau region implies that a relatively low error rate, close to the thermodynamic limit, can be achieved with only a modest cost in terms of futile hydrolysis events. \change{From published values of kinetic parameters, the appropriate discrimination energies for initial unbinding and post-hydrolysis release steps are $\Delta_1 \approx 4, \Delta_2 \approx 8$ (see Appendix A). These values place the system within the plateau region, with the number of excess hydrolyses expected to be quite low (on the order of $\left<n\right> = 0.1$). 
	We note that the error rate corresponding to realistic parameters is roughly an order of magnitude higher than the minimum possible rate, a feature that may arise from the need to optimize for speed as well as energy cost~\cite{banerjee2017elucidating}.}

\begin{figure}
	\centerline{\includegraphics[width=8.3cm]{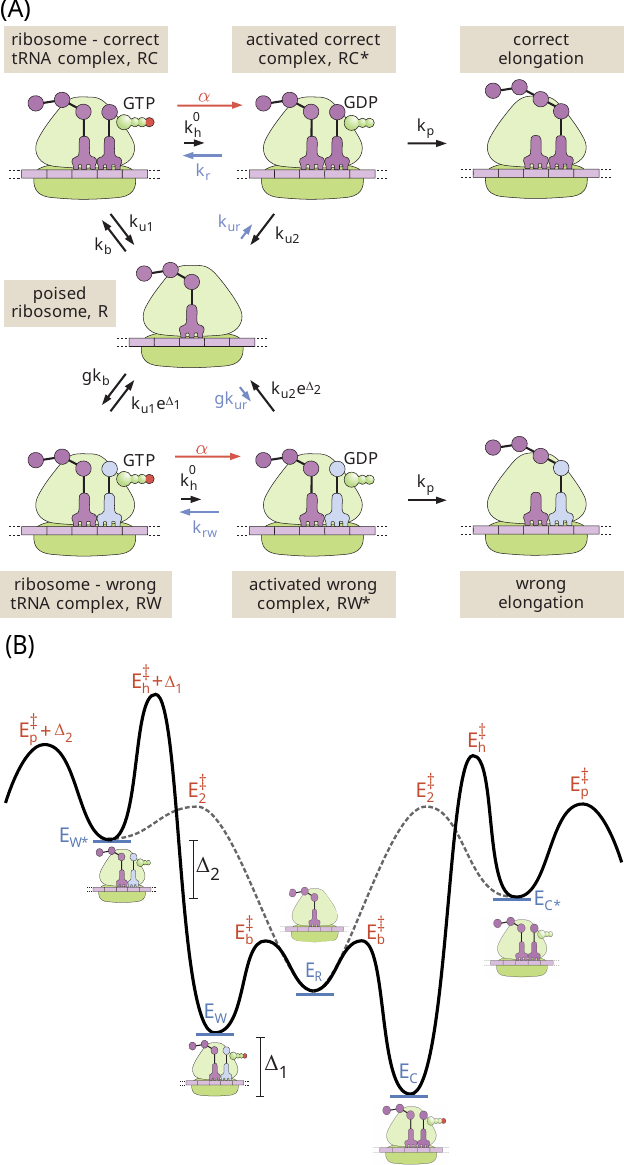}}
	\caption{\textsf{\textbf{Schematic of proofreading model with reversible transitions and thermodynamic driving.}} (a)  Blue arrows indicate reverse transitions not included in the original model. Red arrows show additional driven reaction rate \change{for hydrolysis}. (b) Example energy landscape describing states prior to elongation, in the absence of driving. Energy levels (blue) are shown for a single tRNA, and so do not include the concentration factor $g$. \change{Transition state energies are labeled in red, and are set to give indiscriminate transition rates $k_h^0, k_{ur}$, relegating discrimination factors entirely to unbinding and release. } }
	\label{fig:energylandscape}
\end{figure}

\begin{figure}[h!]
	\centerline{\includegraphics[width=8.3cm]{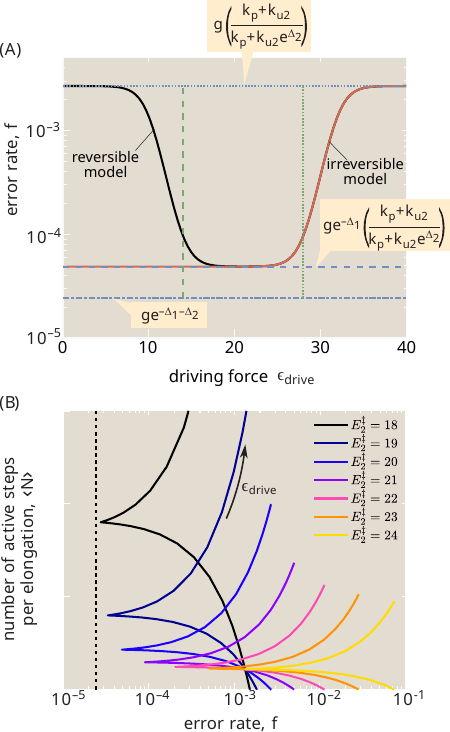}}
	\caption{\textsf{\textbf{Trade-off between accuracy, driving, and cost for reversible proofreading system.}}
		(a) Error rate $f$ is plotted against the thermodynamic driving force $\epsilon_\text{drive}$. Black curve: reversible model. Red curve: irreversible limit with total activation rate set to $k_h = k_h^0e^{\epsilon_\text{drive}}$. Dashed blue line is the limit for the irreversible model with small $k_h$. Dotted blue line is the limit where discrimination occurs entirely in the post-hydrolysis release step. Dash-dotted blue line shows the thermodynamic limit for the minimal possible error rate. The green lines mark the region of optimal driving:
		\change{$E_\text{h}^\ddagger -E_\text{2}^\ddagger+\Delta_1 < \epsilon_\text{drive} < E_\text{h}^\ddagger -E_\text{b}^\ddagger$. Release transition state is set to $E_2^\ddagger = 20$.}
		(b) Average number of active transitions to reach elongation, plotted versus the error rate. Each curve corresponds to a different energy for the transition state $E_2^\ddagger$, with $\epsilon_\text{drive}$ increased along each curve. 
		Dashed black line shows thermodynamic minimal error rate $ge^{-\Delta_1-\Delta_2}$. \change{Parameters are: $\Delta_1 = 4, \Delta_2 = 8, E_h^\ddagger = 30, E_b^\ddagger = 2, E_p^\ddagger = 20, g=4$}. 
	}		
	\label{fig:driveactivation_cost}
\end{figure}

Overall, despite the excess concentration of wrong amino acids, the energetic cost per elongation step is quite low, with \change{little more than one GTP molecule hydrolyzed per new amino acid incorporated}. This is due largely to the fact that the binding discrimination $\Delta_1$ provides an initial non-energy-consuming filter, such that \change{over $90^\%$} of the tRNAs reaching the \change{hydrolyzed} state are already bearing the correct amino acid. %Without this discrimination in the binding energy, the ribosome would need to hydrolyze on average $g=4$ GTP molecules to incorporate each amino acid, in order to achieve the necessary low error rates in translation. 
In the clubhouse analogy, even a rough selection of who comes through the swinging doors can greatly reduce the number of identity checks the demon has to perform.

\subsection{Reversible model: proofreading on a landscape}

\begin{figure}
	\centerline{\includegraphics[width=8.3cm]{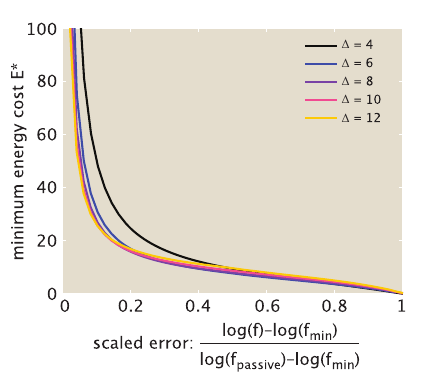}}
	\caption{\textsf{\textbf{Minimal energy cost to reach a desired error rate}}
		Minimal energy per elongation $E^*$ is plotted as a function of the rescaled error rate, defined as an interpolation between the minimum value $f_\text{min} = ge^{-2\Delta}$ and the passive value $f_\text{passive} = ge^{-\Delta}$, on a logarithmic scale.  Both discrimination factors are assumed to be the same, with each curve corresponding to a fixed value of $\Delta_1 = \Delta_2 = \Delta$. \change{The energetic cost is minimized over all values of $E^\ddagger_h, E^\ddagger_2, E^\ddagger_b, E^\ddagger_p$,  with fixed $g =4$.}
	}		
	\label{fig:minenergy}
\end{figure}

The model schematic in Fig.~\ref{fig:scheme} makes use of one-way arrows and thus does not explicitly define the energy cost associated with each hydrolysis cycle. \change{This model is applicable when the splitting probabilities are negligible for the pathway steps reversing the hydrolysis and release processes.}

A more general form of the model with reversible transitions can be used to explore the relation between the thermodynamic driving force and the fidelity of the system. Such a model requires defining an energy landscape for the system, as illustrated in Fig.~\ref{fig:energylandscape}. To constrain the space of possible schemes, we make a few key assumptions. 
First, we assume that the final elongation step, which proceeds at rate $k_p$ is still effectively irreversible. 
The elongated chain thus serves as an absorbing state for the system, and we focus our attention on the kinetics of the transitions preceding this state.

\change{
Because the error rate and number of futile cycles are derived from the splitting probabilities of the system, these quantities are entirely determined by the heights of the transition states on the energy landscape~\cite{banerjee2017accuracy,mallory2020kinetic}. However, in general, transition state energies for reactions in solution tend to correlate with the energies of intermediate and product states~\cite{evans1936further,dill2010molecular,vinu2012unraveling}, implying that we might expect the entire path along the landscape for binding and hydrolysis of the wrong tRNA to be shifted upwards compared to the correct tRNA. In Fig.~\ref{fig:energylandscape}b we present a simplified landscape that incorporates the same assumptions as in our irreversible model (Fig.~\ref{fig:scheme}).
Namely, the bound state energy is assumed to be higher by $\Delta_1$, and the hydrolyzed state higher by $\Delta_2$ for the wrong versus the correct tRNA. The discrimination is again taken to be in the release rates ($k_{u1},k_{u2}$) only, while the forward hydrolysis rate ($k_{h0}$) and the elongation rate ($k_p$) are assumed the same for both. The binding rate $k_b$ and reverse-release rate $k_{ur}$ are taken to differ only by the concentration factor $g$ describing the increased likelihood of wrong tRNA binding. In the resulting energy landscape, these assumptions imply that the transition state for hydrolysis is higher by $\Delta_1$  the transition state for elongation is higher by $\Delta_2$ for the wrong tRNA. While other assumptions for the energy landscape shape are possible~\cite{banerjee2017accuracy}, this approach yields a thermodynamically consistent model that directly generalizes the classic irreversible system discussed in the previous section.}

\change{The high energy of the intermediate states $\text{RC}^*, \text{RW}^*$ allows the peptide bond to be kinetically trapped, with the elongation step being effectively irrevesible.
As a consequence, the equilibrium probability of those states and the concomitant elongation flux must be very low.} An additional driven process is needed to push the system towards elongation (red arrow in Fig.~\ref{fig:energylandscape}a) . This process could represent the hydrolysis of GTP and/or release of GDP from the Ef-Tu elongation factor. We assume the rate $\alpha$ associated with this driving is the same regardless of the tRNA identity. The corresponding thermodynamic driving force can then be expressed as $\epsilon_\text{drive} = \log(1 + \alpha/k_{h}^0)$~\cite{lin2020circuit, arunachalam}.

The model with reverse transitions can be solved as before by combining the  splitting probabilities to compute the error rate $f$ of elongating with the wrong versus the correct tRNA, and the  average number of transitions through the driven hydrolysis step  to reach elongation ($\left<N\right>$). The final expressions (with derivation provided in Appendix B) are:
\begin{align}
f  & =  \frac{p_{pw}(p_{bw} p_{hw} + p_{urw})/(1-p_{rw}p_{hw}) }{p_{pc}(p_{bc} p_{hc} + p_{urc})/(1-p_{rc}p_{hc})}, \label{eq:f_reversible}\\
\left<N\right> & =  \left[\frac{(\frac{p_{bc} + p_{urc}p_{rc})p_{hc}}{1-p_{rc}p_{hc}} + \frac{(p_{bw} + p_{urw}p_{rw})p_{hw}}{1-p_{rw}p_{hw}}}
{
\frac{(p_{bc}p_{hc} + p_{urc})p_{pc}}{1-p_{rc}p_{hc}} + \frac{(p_{bw}p_{hw} + p_{urw})p_{pw}}{1-p_{rw}p_{hw}}
}\right](1-e^{-\epsilon_\text{drive}})
\label{eq:N_reversible}
\end{align}
where $p_{urc},p_{urw}$ are the splitting probabilities for going directly from the empty $R$ state into the \change{hydrolyzed} state along the reverse release pathway, $p_{rc}, p_{rw}$ are splitting probabilities for the reverse \change{hydrolysis} transition, and $p_{hc}, p_{hw}$ are splitting probabilities of going towards the \change{hydrolyzed state} from the bound state (along either the basal or the active arrow).

As plotted in Fig.~\ref{fig:driveactivation_cost}A, increasing the driving force for activation has a non-monotonic effect on the accuracy of the system. When there is no driving, the barrier $E_h^\ddagger$ is so high that the system primarily reaches the RC$^*$/RW$^*$ state through the reverse-release pathway (rates $k_{ur}, gk_{ur}$, \change{gray curve in Fig.~\ref{fig:energylandscape}B}). In this case, the difference in binding energy $\Delta_1$ becomes irrelevant and the error rate approaches the known value for a simple substrate-selective enzymatic reaction~\cite{johansson2008rate,tawfik2014accuracy}:
\begin{equation}
f\rightarrow g \frac{k_p + k_{u2}}{k_p + k_{u2} e^{\Delta_2}}.
\label{eq:fidelity_nobind}
\end{equation}
As the driving force rises, the binding and activation pathway \change{(black curve in Fig.~\ref{fig:energylandscape}B)} begins to dominate, and the error rate decreases until the \change{hydrolysis} transition becomes effectively irreversible. The only splitting probabilities in Eq.~\ref{eq:f_reversible},~\ref{eq:N_reversible} that depend on the driving force are $p_{hc},p_{hw}$. The transition to the irreversible system occurs when $p_{bw}p_{hw} > p_{urw}$, or equivalently when \change{$\epsilon_\text{drive} > E_\text{h}^\ddagger -E_\text{2}^\ddagger+\Delta_1$}.
At that point, the error rate reaches the value given in Eq.~\ref{eq:f_smallh} for the irreversible system, and further driving does not improve the accuracy. If the driving force becomes much higher, the rapid \change{hydrolysis} transition prevents the system from sensing the difference in binding energy $\Delta_1$ and it again approaches the limit in Eq.~\ref{eq:fidelity_nobind} where the only discriminating step is release from the hydrolyzed state. The transition to this increased error rate occurs when $k_h^0 + \alpha > k_{u1}$ or equivalently when \change{$\epsilon_\text{drive} > E_\text{h}^\ddagger -E_\text{b}^\ddagger$}.

In the reversible model, the energetic cost for elongating the peptide by one amino acid can be expressed as $E^* = \left<N\right> \epsilon_\text{drive}$: the product of the thermodynamic driving force for each activation cycle and the number of such cycles required per elongation event. The tradeoff between futile cycles and accuracy is evident in Fig.~\ref{fig:driveactivation_cost}b. \change{Approaching closer to the minimal possible error rate requires a lower barrier $E_2^\ddagger$, which in turn enables more frequent release events, thereby raising the number of futile cycles and the energy cost.}

\begin{figure}[t]
	\centerline{\includegraphics[width=8.3cm]{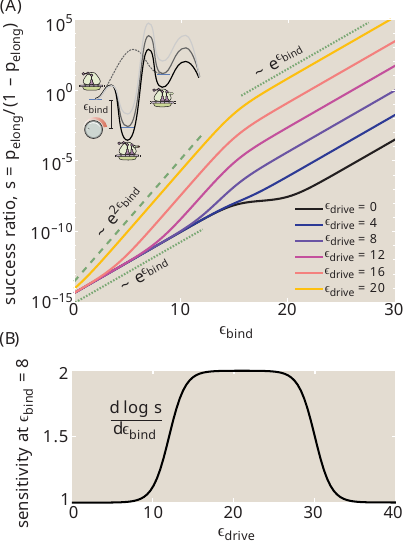}}
	\caption{\textsf{\textbf{Energetic driving enhances sensitivity to substrate binding}} (a) For a single type of tRNA ($g=0$), probability that an interaction will successfully result in elongation is plotted against the binding energy $\epsilon_\text{bind}$. The energy landscape along the binding-and-activation pathway is shifted concomitantly with the binding energy, maintaining constant barriers for transitions (inset). 
	Each curve corresponds to a different driving force $\epsilon_\text{drive}$. Dashed and dotted green lines show two different exponential scalings. (b) Sensitivity to the binding energy (defined as $d\log p_\text{elong}/d\epsilon_\text{bind}$), at $\epsilon_\text{bind}=8$ is plotted as a function of driving force. Parameters used: $E_2^\ddagger = 20, E_h^\ddagger+\epsilon_\text{bind} = 40, E_p^\ddagger+\epsilon_\text{bind} = 35$.}
	\label{fig:sensitivity}
\end{figure}

 The range of possible error rates, and the total energy $E^*$ needed to achieve a certain error, depend on the binding energy differences $\Delta_1, \Delta_2$ distinguishing correct versus wrong tRNAs. The minimal possible error for the active system is given by $f_\text{min} = g e^{-\Delta_1-\Delta_2}$. A passive equilibrium system can only achieve the error of $f_\text{passive} = g e^{-\max(\Delta_1,\Delta_2)}$.  By numerically minimizing \change{over the transition state heights}, we can compute the minimal cost for sliding between these two error limits, as shown in Fig.~\ref{fig:minenergy}. Errors above $f_\text{passive}$ can be achieved at zero cost. Pushing towards the minimal possible value of $f_\text{min}$ requires an infinite energetic cost. For reasonable values of the binding energy difference $\Delta$ (corresponding to a few hydrogen bonds), an energy cost on the order of $10-20k_bT$ per elongation step is sufficient for approaching close to the minimal error, after which the cost begins to grow steeply.

Energy dissipation in the translational proofreading system allows it to more accurately discriminate among tRNAs with similar binding energies. This property can be couched in terms of sensitivity or signal gain: the input signal is the binding energy of a particular tRNA ($\epsilon_\text{bind}$) and the output is the likelihood that the tRNA will successfully transfer its amino acid to the peptide chain each time it interacts with the ribosome. In Fig.~\ref{fig:sensitivity}, we quantify this output by plotting the success ratio $s = p_\text{elong}/(1-p_\text{elong})$ for a system with only one type of tRNA present ({\em i.e.} $g=0$). The more sensitive this ratio is to the binding energy, the more capable the system will be of distinguishing between tRNAs with small binding energy differences. \change{Note that we assume transition state heights are tied to the intermediate bound state (Fig.~\ref{fig:sensitivity}A, inset), so that the hydrolysis and elongation rates ($k_{h0},k_{p}$) remain constant throughout.}

For a passive system with $\epsilon_\text{drive}=0$, the success ratio scales exponentially with the binding strength in both the strong-binding and weak-binding limits (Fig.~\ref{fig:sensitivity}, black curve). In the presence of strong energetic driving and weak binding, the success ratio exhibits a quadratically steeper scaling, implying greater sensitivity of the system. When binding is very strong, then unbinding becomes vanishingly unlikely and the system loses its ability to stack multiple binding energy differences, reverting back to the lower sensitivity. The classic definition of sensitivity as the derivative of the logarithm of the output~\cite{owen2023size} is plotted in Fig.~\ref{fig:sensitivity}B, demonstrating that intermediate driving confers the greatest sensitivity values.

\begin{figure*}
	\centerline{\includegraphics[width=15.5cm]{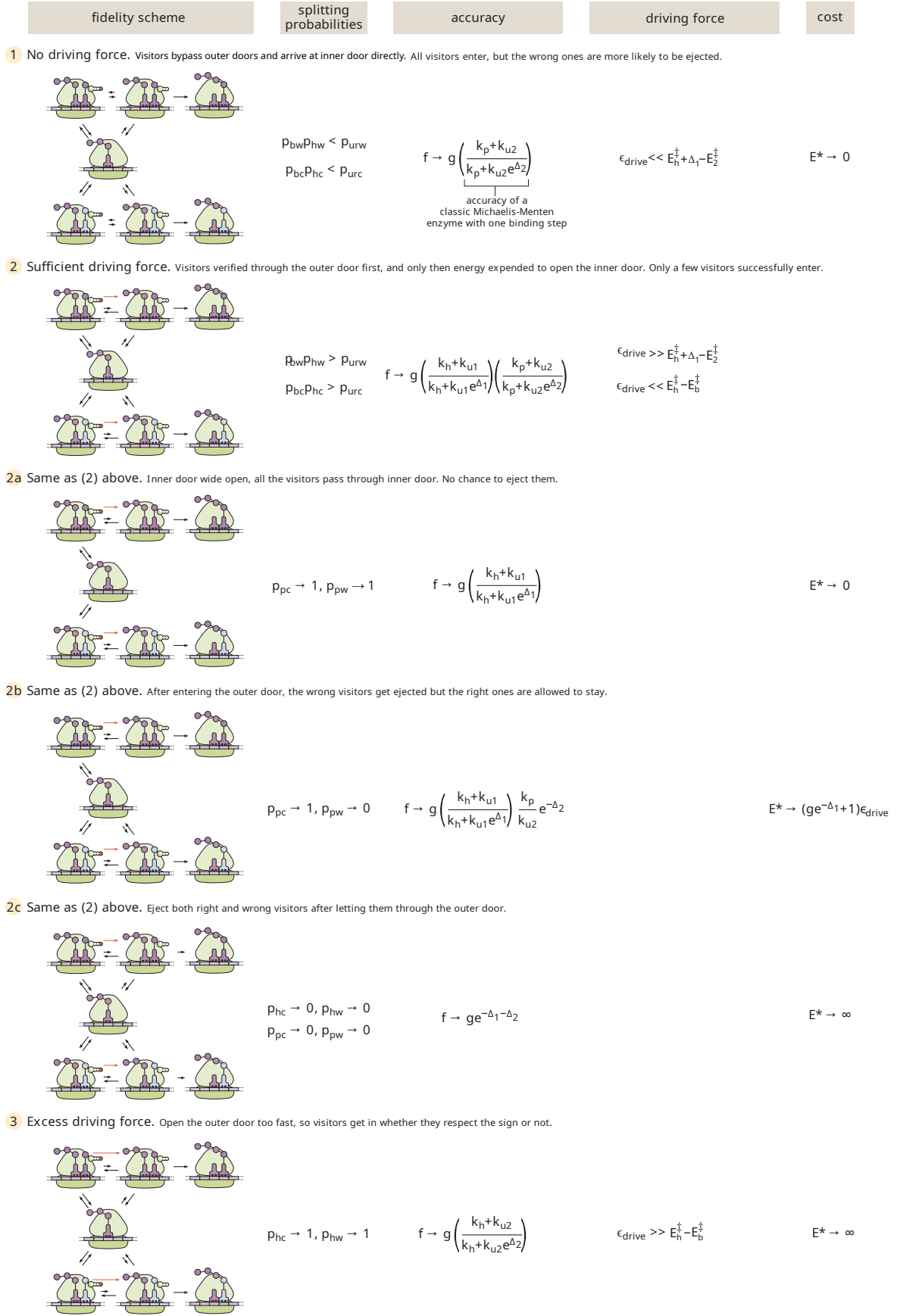}}
	\caption{\textsf{\textbf{Summary of energy cost and accuracy for different regimes  in the translational proofreading model.}} Limiting cases are shown for (1) No driving, (2) Sufficient driving to approach the irreversible model, and (3) Excess driving. The splitting probability limits, the error rate $f$, the driving force $\epsilon_\text{drive}$, and the total energetic cost $E^*$ are listed for each regime. }
	\label{fig:regimes}
\end{figure*}

As summarized in Fig.~\ref{fig:regimes} this simple proofreading system exhibits three regimes with increasing driving force, which can be seen in both Fig.~\ref{fig:driveactivation_cost}A and Fig.~\ref{fig:sensitivity}B. At very low driving, hydrolysis is extremely unlikely, and elongation can only be achieved when the tRNA bypasses the binding and hydrolysis step to enter the active state directly. \change{The error rate is then determined entirely by the discrimination in the post-hydrolysis release ($\Delta_2$), corresponding to a sensitivity of $1$.}
At intermediate driving, the hydrolysis pathway dominates, and the irreversible model becomes an adequate description of the system. Within this regime, the error rate is determined multiplicatively by two factors that each correspond to the error rate of a single Michaelis-Menten enzyme. Each factor involves a balance between the release rates and the rate of transitioning forward to the next state. In the limit where release dominates, the error rate scales exponentially with the sum of both discrimination energies ($\Delta_1+\Delta_2$), but the requisite number of futile cycles and the concomitant energetic cost approaches infinity.
This limit corresponds to a sensitivity of $2$. 
When excess driving is applied, the tRNA has no chance to unbind before hydrolysis and the error rate is again determined only by a single discrimination energy ($\Delta_2$), with sensitivity approaching $1$. 

\change{We note that the parameters $\Delta_1, \Delta_2$ are canonically thought of as binding energy differences for the wrong versus correct tRNA in the initially bound and in the high-energy intermediate state, respectively~\cite{hopfield1974kinetic,murugan2012speed}. However, they can also be more generally described in terms of the transition state heights and in terms of their effect on the splitting probabilities for unbinding or release of the tRNAs.}

Overall, translational proofreading serves as an illustrative case study of the trade-offs between energetic driving, total cost of futile cycles, and the accuracy of the system in distinguishing between substrates with small differences in binding energy.

\section{Catalytic control of dynamic instability}
 
In the simple proofreading scheme described above, active driving enables the system to better differentiate the right vs wrong substrate, despite the fact that the driven hydrolysis rate itself is \change{taken to be independent of the substrate}. Proofreading thus makes the system more sensitive to the pre-existing differences in release rates for the two substrates. We proceed to further explore this feature of exploratory dynamics with resetting: the ability to regulate system behavior by tuning passive transition rates.

Specifically we focus on control by reusable catalysts present in much smaller quantities than the reactants themselves.
At thermodynamic equilibrium, the presence of a catalyst can only alter the transition barriers and thus shorten the relaxation to the steady state but not change the steady-state distribution itself. However, many biomolecular systems engage in catalytic control, where a catalyst modulates the probability distribution at steady state. Unlike allosteric control via ligand binding~\cite{phillips2020molecular}, catalytic control implies that the regulator molecule can be reused over and over again, while the substrates maintain some memory of their interaction with it. Consequently, the catalyst can be present at sub-stoichiometric concentrations compared to the substrate. For example, some kinases are more than three orders of magnitude lower in concentration than their target substrates \cite{kinase_concentration}. The prevalence of catalytic control alleviates spatial crowding constraints when multiple regulatory proteins are necessary to tune protein activity or assembly.

Catalytic control requires the system to be driven out of equilibrium. Recent work demonstrated that the change in probability distribution due to the addition of a catalyst is bounded by twice the total applied thermodynamic force~\cite{arunachalam}.
The presence of driving can make the steady state of a system sensitive to catalysts, facilitating other transitions that are not themselves actively driven. Such systems are qualitatively distinct from molecular machines that dissipate energy to directly push the system towards a target state. Instead, they rely on exploratory dynamics that sample many pathways, enabling multiple points of regulation by different catalysts present in sub-stoichiometric concentrations.

\begin{figure}
	\centerline{\includegraphics[width=8.3cm]{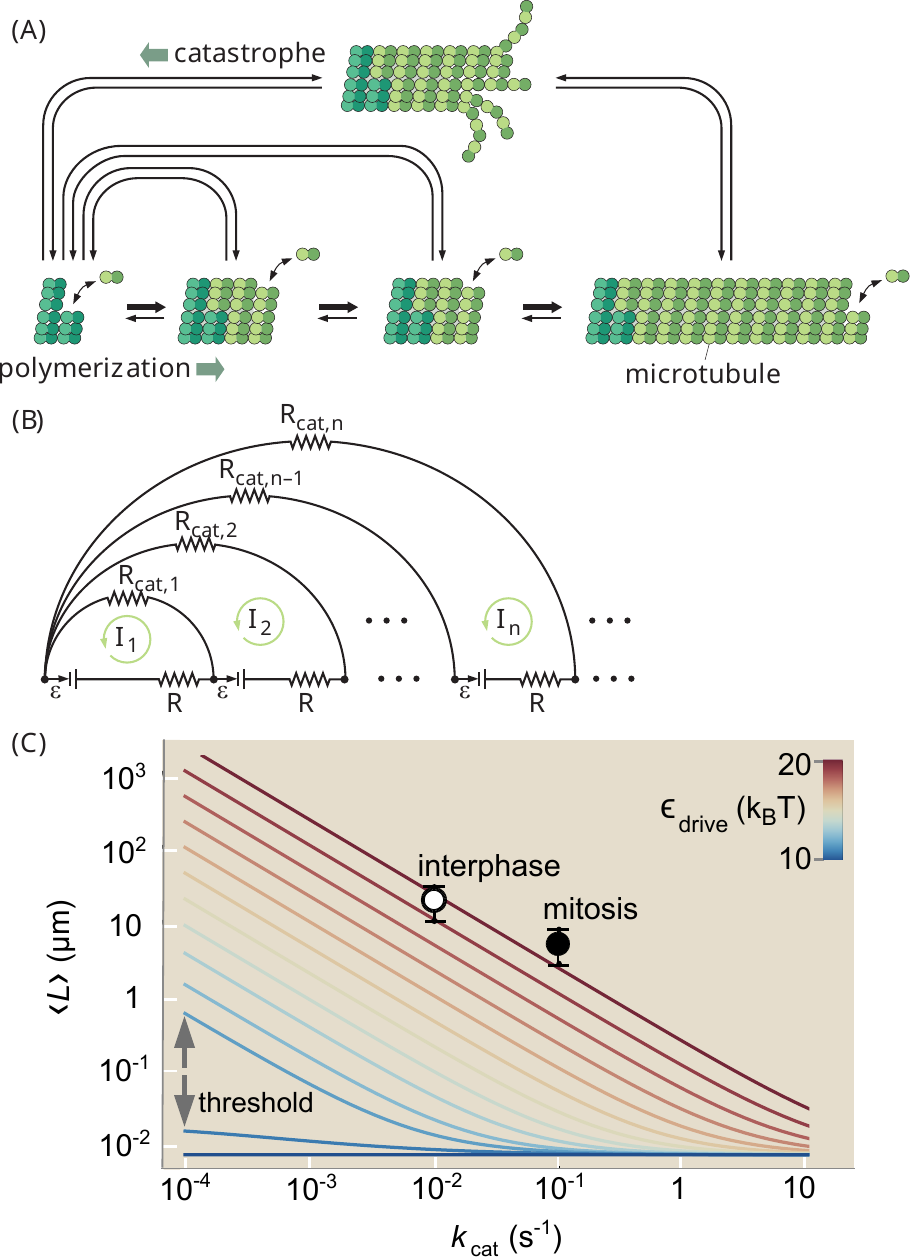}}
	\caption{\textsf{\textbf{Catalytic regulation of microtubule length.} Self-assembly of tubulin subunits into filaments, with complete dis-assembly events (catastrophes) (A). This process can be mapped to a circuit diagram (B). Eq. 1 gives the mean length as a function of catastrophe rate and thermodynamic force, which can be interpreted as the GTP concentration (color bar; C). At equilibrium (blue line), catastrophe rate has no influence on mean length. At physiological GTP concentration (red), the predicted $k_\text{cat}$ dependence is in excellent agreement with the measured mean microtubule length at interphase and mitosis, which differ only in the catastrophe rate (C). The catastrophe rate linearly tunes mean microtubule length above a critical thermodynamic force (GTP concentration). Parameters and measured lengths are taken from {Refs.\cite{belmont1990real}} and \cite{microtubule_howard}. }
	}
	\label{fig:catalytic}
\end{figure}

\subsection{Microtubule length control model}
Here, we illustrate a concrete manifestation of this phenomenon in the context of microtubule length regulation by catalytic factors that destabilize the end cap at the tip of the growing microtubule.
\change{As with the translational proofreading case, our goal is to reduce the system to the simplest possible model that encompasses the features necessary for catalytic control. 
	The average microtubule length is regulated by the cell, and is known to vary at different points in the cell cycle~\cite{belmont1990real}. The minimal model described below demonstrates how the presence of driving in the assembly step enables the average length to be sensitive to changes in the rate of catastrophe, which is the only microtubule-associated rate constant that is upregulated by the cell upon entry into mitosis \cite{belmont1990real}. }
	
 The elementary steps constituting microtubule self-assembly are shown in Fig.~\ref{fig:catalytic}A in the absence of rescue from catastrophe \cite{dynamic_instability}. The state space contains an infinite number of possible states corresponding to increasing lengths of the microtubule.
The forward elongation process in this simple model is followed by the hydrolysis of GTP to GDP, whose rate determines the size of the  GTP-containing microtubule end-cap.

In a system where GTP and GDP are allowed to fully equilibrate, the reversible assembly of tubulin onto the filament occurs with forward and reverse rate constants $k_f$ and $k_r$.
 Thus, $k_f/k_r=e^{-\beta G}$, where $G$ is the equilibrium dimer binding free energy. 
In cells, GTP is kept at high concentration in excess of its equilibrium level, effectively giving rise to an additional forward rate constant $\alpha$, which is proportional to excess [GTP] up to a saturation concentration. The thermodynamic driving force is $\epsilon_\text{drive} =  \ln{[1+\alpha/k_f]}$. 

Assembly is counteracted by a catalyzed catastrophe process with rate constant $k_{\mathrm{cat}}$, allowing for the complete disassembly of the microtubule in a regulatable manner \cite{luke2,bowne2015regulation}. 
Catastrophe is triggered by the stochastic disruption of the growing microtubule cap \cite{microtubule_capping}, which allows cap-modifying substrates to act as sub-stoichimetric catalysts of microtubule shrinkage. Microtubule-associated proteins that trigger catastrophe include both ATP-consuming motors in the kinesin family~\cite{walczak2013microtubule} and passive factors such as Op18/Stathmin~\cite{belmont1996identification}. Notably, stathmin levels in the cell are estimated to be sub-micromolar~\cite{ringhoff2009stathmin}, while tubulin can reach concentrations in the hundreds of micromolar~\cite{Baumgart2019}, indicating that this enzyme must act as a reusable catalyst.

Microtubule catastrophe serves as a resetting step for the exploratory dynamics of the microtubule length. 
 The dynamic instability steady state is reached when catastrophe balances net dimer addition, resulting in a length distribution that is distinct from the equilibrium steady state of the system~\cite{dynamic_instability} and is dependent on the level of catalysist present.

 Although these processes have been modeled mathematically~\cite{microtubule_langevin} and via computational simulations~\cite{microtubule_simulations}, the complexity of the dynamical system consisting of numerous reversible reactions, has limited our quantitative understanding of how system parameters control microtubule length distributions. 
 Previous work has established the intrinsic speed-up of non-equilibrium polymer reorganization kinetics compared to equilibrium reorganization~\cite{nonequilibrium_polymerspeed}. An article within the current issue~\cite{Kondev2025} highlights how the exploratory dynamics of growing microtubules enable them to rapidly find targets within the cell. The resetting catastrophe process thus provides clear benefits to the speed of the system. Here we highlight the additional advantage of non-equilibrium driving in enabling steady-state length regulation via a catalyst.

\subsection{Length distribution depends on catalytic rate}
Equilibrium theory teaches that catalytic rate constants cannot affect the mean value of any observable. In contrast, the microtubule length probability distribution $P(L)$ reaches a steady-state where the mean length is known to depend explicitly on the catalytic rate $k_{\mathrm{cat}}$. In the limit of strong driving, with near-irreversible catastrophe and forward-biased growth, the mean length has previously been computed as $\left<L\right> = (\alpha+k_{\mathrm{f}}-k_{\mathrm{r}})/k_{\mathrm{cat}}$ \cite{microtubule_leibler}.
In this regime, a catalyst which only decreases the energy barrier to catastrophe leads to a proportional change in the mean length, in violation of the equilibrium rule. Such catalytic regulation is known to occur during the eukaryotic cell cycle, where increased $k_{\mathrm{cat}}$ causes the decrease in microtubule length necessary for cell division \cite{belmont1990real}. However, the switching on of catalytic control as a function of thermodynamic driving has not been established.

The reaction scheme in Fig.~\ref{fig:catalytic}A is a generalization of the single-step proofreading circuit shown in Fig.~\ref{fig:scheme}. In the appendix, we demonstrate how the method of counting weighted paths can be applied to a simplified system where the catastrophe process is irreversible. We use this approach to compute the distribution of lengths at which catastrophe occurs. For catastrophe to be nearly irreversible in an equilibrium system, the free energies of longer states must be much higher, and most microtubules would only reach a very short length before undergoing catastrophe. For a driven system, however, it is possible to extend this distribution to arbitrarily long lengths by raising the probability of stepping forward rather than reversing or undergoing catastrophe at each state. Because this probability depends on the catastrophe rate $k_\text{cat}$, the resulting system is necessarily sensitive to the level of catalytic enzyme.
This sensitivity is analogous to proofreading fidelity, allowing the system to accurately convert different levels of catalyst to different responses. 

We note that the path-counting approach becomes prohibitively tedious for complex reaction systems, including when reverse catastrophe transitions are included.  The approach can be automated in the form of matrix algebra~\cite{scott2021diffusive}, or replaced with approaches that rely on solving the chemical master equation~\cite{yu2022energy} or on graph-theoretic methods that count spanning trees across the network~\cite{wong2020gene}. However, to maximize our intuition regarding the role of energetic driving, we turn to an alternate technique that involves mapping the system to an effective circuit framework~\cite{lin2020circuit} (Fig.~\ref{fig:catalytic}B), with batteries representing driven transitions ($\mathcal{E} \propto \alpha/k_f$). 

By leveraging techniques for simplifying electronic circuits, we can then compute a  closed-form expression for the steady-state length distribution of microtubules (See Appendix C):
\begin{multline}
P(L)/P(1)= {{k_{\mathrm{cat}}}\over{k_{\mathrm{cat}}-\alpha (e^{\beta G}-1)}}e^{-\beta G(L-1)} \\\\ +{{\alpha(e^{\beta G}-1)}\over{\alpha (e^{\beta G}-1)-k_{\mathrm{cat}}}} e^{-D(L-1)},
\label{eq:microlength}
\end{multline}
where $P(1)$ is the monomer fraction, and $D \equiv - \ln{[1-{{\sqrt{(\alpha+k_{\mathrm{cat}}+k_f-k_r)^2+4k_{\mathrm{cat}}k_r}-(\alpha+k_{\mathrm{cat}}+k_f-k_r)}\over{2k_r}}]}$. When catastrophe is equally likely from all states, this expression can also be used to compute the distribution of lengths at which catalysis occurs: $P_\text{cat}(L) = P(L)/[1-P(1)]$, matching the results obtained by counting paths. 

Although mean filament length has been calculated using generating functions \cite{microtubule_length}, this is the first time that the full distribution $P(L)$ has been solved and the role of the thermodynamic force isolated. Interestingly, $P(L)$ is  a superposition of two exponential functions, corresponding to the equilibrium and nonequilibrium contributions, respectively. The double exponential explains why previous attempts to fit $P(L)$ generated from numerical simulations to a single exponential distribution led to poor fits~\cite{microtubule_length}.  

Fig.~\ref{fig:catalytic}C shows the mean microtubule length as a function of catastrophe rate as predicted by Eq.~\ref{eq:microlength}, using measured rate constants \cite{microtubule_howard}, for varying $\alpha$ corresponding to different GTP concentrations. As expected, if $\alpha = 0$ (blue line) then Eq.~\ref{eq:microlength} reduces to the equilibrium single-exponential distribution, which is independent of $k_{\mathrm{cat}}$. However, as the system is driven from equilibrium, the length distribution jumps between two distinct regimes with qualitatively different dependence on $k_{\mathrm{cat}}$. The jump occurs when $\alpha$ exceeds $k_r - k_f$. In the strongly-driven regime, for which $(\alpha+k_{\mathrm{f}}-k_r)/k_{\mathrm{cat}} \gg 1$, Eq.~\ref{eq:microlength} simplifies to $\left<L\right>_{\mathrm{strong}} = (\alpha+k_{\mathrm{f}}-k_{\mathrm{r}})/k_{\mathrm{cat}}$, which is the well-known formula cited above. At physiological GTP concentrations, the predicted mean length is in excellent agreement with measured lengths \cite{belmont1990real} in both mitosis and interphase (circles in Fig.~\ref{fig:catalytic}C). In the weakly-driven regime ($-(\alpha+k_{\mathrm{f}}-k_r)/k_{\mathrm{cat}} \gg 1$), Eq.~\ref{eq:microlength} simplifies to $\left<L\right>_{\mathrm{weak}}=  -\ln{[{{\alpha+k_f}\over{kr}}+{{k_{\mathrm{cat}}(\alpha+k_f)}\over{k_r(\alpha+k_f-k_r)}}]}^{-1}$; the mean length is only marginally sensitive to $k_{\mathrm{cat}}$ in this regime. The thermodynamic force, as parameterized by $\alpha$ or [GTP], controls the transition between the near and far-from-equilibrium regimes, whose sharpness is inversely proportional to $k_{\mathrm{cat}}$ (Fig.~\ref{fig:catalytic}C). Therefore, a uniquely non-equilibrium feature (catalytic regulation of an ensemble-averaged observable) is turned on in a switch-like manner when the system is driven beyond the threshold level. 

\section{Discussion}
In this work we highlight how exploratory dynamics with resetting enhances the sensitivity of biochemical pathways. In the case of translational proofreading, release from a high-energy intermediate state increases the ability of the system to select the correct tRNA among an excess of decoys, rendering it more sensitive to  small differences in binding energies. For the case of microtubule length control, resetting through catastrophe allows the system to be responsive to sub-stoichiometric concentrations of a destabilizing enzyme. In both cases, sensitivity comes at an energetic cost, requiring GTP hydrolysis to drive the resetting cycles.

We compute the energetic cost associated with translational proofreading by starting with the classic Hopfield model, which assumes a single GTP is hydrolyzed at each irreversible activation step. 
An intuitively simple probabilistic  approach gives an expression for the number of excess activation cycles. The resulting total energetic cost increases monotonically with the release rate from the \change{hydrolyzed} state, while the error rate of the system decreases. Notably, there is a broad plateau region for intermediate release rates where the number of excess activation cycles is well-approximated by the equilibrium error rate for the initial binding step: $\left<n\right> \approx g e^{-\Delta_1}$. The plateau spans the parameter regime where most correct tRNAs proceed towards elongation while most wrong ones are released. In this case, futile activation cycles occur only when the wrong tRNA passes through to the \change{hydrolysis} step. 

A key consequence of this plateau is that, despite the excess of wrong tRNAs, the energetic cost for elongation remains quite low. The system capitalizes on \change{passive} discrimination during the initial binding and codon recognition to limit the frequency of wrong tRNAs proceeding through \change{hydrolysis}. Since correct tRNAs are less likely to be released, this means that only a small number of GTP hydrolysis events are needed per incorporated amino acid. Conceptually this is akin to letting visitors self-filter through a passive set of swinging doors before allowing them to proceed to an energy-consuming identity check. Partially accurate discrimination in the first step implies that only a few identity checks are needed before an acceptable visitor is permitted to enter.

Models with irreversible transitions are, in principle, unphysical, requiring an infinite input of energy to completely preclude reverse transitions. In practice, however, such models are meant to represent schemes where the reverse transition is so unlikely that it does not contribute to the splitting probabilities of the system. As shown for the translational proofreading example, increasing energetic driving can push a system towards the irreversible limit; further driving beyond that necessary to reach this limit can actually hinder the sensitivity of the system. Thus, irreversible models constitute a useful limiting case for quantifying the efficiency of a system undergoing exploratory dynamics.

\change{The minimalist translational proofreading model analyzed here makes several simplifying assumptions -- notably, that discrimination of wrong versus right tRNAs is localized to the release steps only. By contrast, experimental measurements of ribosome kinetics indicate different kinetic rates for the forward hydrolysis and elongation transitions as well~\cite{rodnina2001ribosome,wohlgemuth2011evolutionary}. Our mathematical approach, which relies on computing probabilities of different resolutions for each individual ribosome-tRNA interaction event, can be applied to more complex mechanisms by coarse-graining reaction networks as described in Appendix A. Furthermore, we show in the appendix how realistic kinetic models with many discriminating steps can be reduced to the simple system  through the selection of the appropriate discrimination factors $\Delta_1, \Delta_2$ which encompass the ability of the system to distinguish between wrong and right substrates before and after the active hydrolysis step.
	
	 While the interplay of speed, energy dissipation, and fidelity for ribosomal translation have been extensively explored in prior work~\cite{murugan2012speed,banerjee2017elucidating,yu2022energy}, this manuscript provides a pedagogically simple `blackboard-friendly' approach for calculating the energetic cost of accuracy in the presence of an excess concentration of wrong tRNAs. These calculations enable an intuitively clear picture of the necessary parameter regime for accurate and energy-efficient translation: one where a moderately low error rate leading to the hydrolysis step is driven down still further by the proofreading process. Furthermore, we show the necessary bounds on energetic driving, which must be large enough to approach an irreversible limit yet not so large as to bypass the passive discrimination step.}

In kinetic proofreading, energy dissipation at one point in the system increases the sensitivity to small differences in release rates elsewhere, enabling accurate discrimination in the presence of excess decoys. The sensitivity jumps sharply when the thermodynamic driving force exceeds a certain critical value. Analogously, we showed that the sensitivity of mean microtubule length to the rate of catastrophe undergoes a switch from logarithmic dependence near equilibrium to linear dependence far from equilibrium. Energetic driving of systems undergoing exploratory dynamics can thus trigger a qualitative transition in their input-output response functions.

 Overall, our results parallel past work~\cite{murugan2012speed,banerjee2017elucidating,yu2022energy} linking the \change{accuracy and energy efficiency} of active proofreading systems. We focus on the specific cases of translational proofreading and catalytic control to concretely illustrate biochemical systems that face a trade-off in energetic cost versus function. Two pedagogically useful approaches are demonstrated to analyze these systems: probabilistic path-counting  and mapping to an electrical circuit system. Both case studies 
 highlight the importance of  driven resetting steps for enhancing sensitivity in conditions where decoy substrate concentrations may be high or regulator concentrations are limited.

 We hypothesize that many other examples of intracellular exploratory dynamics, including quality control pathways, signaling cascades, cell-cycle associated transitions, and organelle rearrangements may be analyzed in an analogous manner to link the energetic cost with the sensitivity to various control parameters. Future exploration of such systems may help elucidate the many functional consequences of the homeostatic energy consumption that defines living cells.\\

\section{Acknowledgements}
 We are grateful to the CZI Theory Institute Without Walls for supporting our joint efforts. 
 RP is grateful to the NIH for support through award numbers DP1OD000217 (Director's Pioneer Award) and NIH MIRA 1R35 GM118043-01. MML is grateful for support from the Sloan Foundation Matter-to-Life grant G-2024-22449 and the NIH R01GM125748. EFK acknowledges support from NSF grant 2034482 and from a UCSD Academic Senate research grant.
 
 \section{Declaration of Interests}
 EFK is a member of the Biophysical Journal editorial board.
 
 \section{Author contributions}

All authors conceived and designed the research, wrote the manuscript, and created figures. EFK carried out the mathematical development for the translational proofreading model and MML did so for the catalytic control model. 

\bibliography{translationEnergyCost.bib}

	\clearpage
	\appendix
	
	\onecolumngrid
	
\section{Appendix A: Coarse-graining realistic proofreading schemes}
	
\begin{figure}
		\centerline{\includegraphics[width=\textwidth]{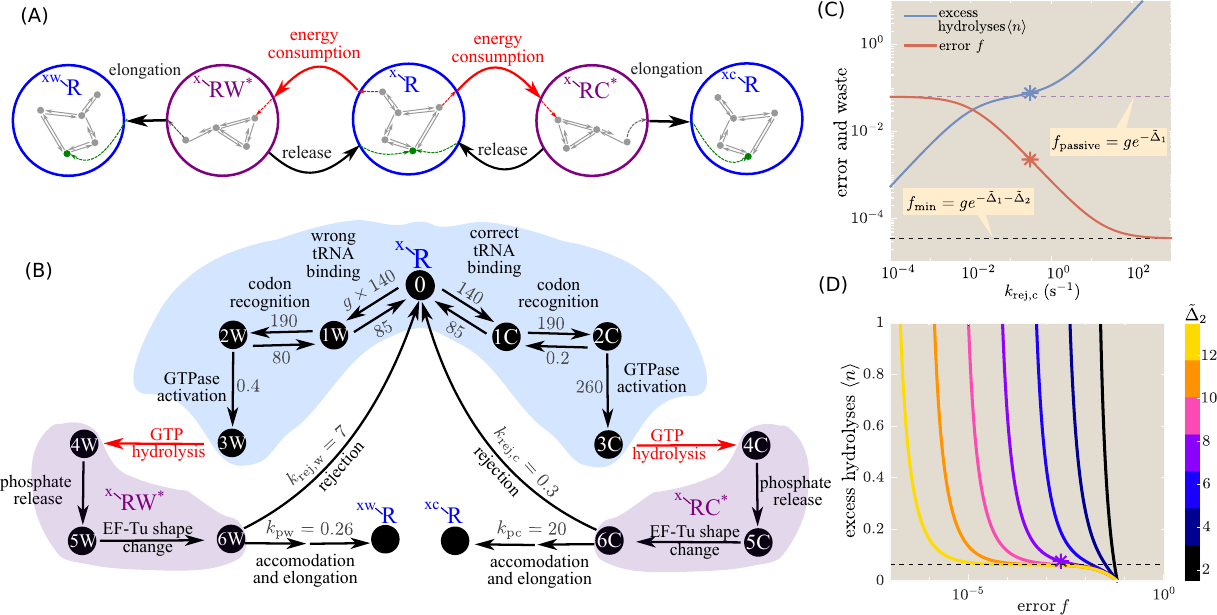}}
		\caption{\textsf{\textbf{Alternate schemes for translational proofreading with a single energy-consuming step.}} (A) General schematic, where each circle corresponds to a compound state containing no internal driven transitions. Red arrows represent the only energy-consuming steps in the system. \change{(B) Example of more complex proofreading scheme, employed in Ref.~\cite{rodnina2001ribosome,wohlgemuth2011evolutionary,zaher2010hyperaccurate} to describe experimental measurements of ribosomal translation. 
		Shaded regions delineate coarse-graining to the compound states in (A). Rate constants (gray) from Ref.~\cite{wohlgemuth2011evolutionary}. All rate constant units are in $\text{sec}^{-1}$, except for the binding rate in $\mu\text{M}^{-1}\text{s}^{-1}$.  %Blue: initial compound state, purple: activated compound state.
		(C) Plot of error $f$ and energetic cost in terms of excess hydrolyses $\left<n\right>$, for the realistic ribosomal translation model and parameters given in (B), with varying rejection rates. The
		rejection rate for the wrong pathway is set to $k_\text{rej,w} = k_\text{rej,c}e^{\tilde{\Delta}_2}$, with the discrimination factor $\tilde{\Delta}_2=7.5$, and all other rates are kept constant throughout. Dashed lines indicate limits for error rate with slow and rapid rejection. (D). Plot of excess hydrolysis cost $\left<n\right>$ versus the error rate $f$, for different values of the proofreading discrimination factor $\tilde{\Delta}_2$, using the realistic model and parameters from (B). Dashed line marks the plateau region $\left<n\right> \approx  g e^{-\tilde{\Delta}_1}$. Stars in (C) and (D) mark the specific measured parameters from Ref.~\cite{wohlgemuth2011evolutionary}.}
		}
		\label{fig:genscheme}
	\end{figure}

	The probabilistic approach described here can be generalized to  more complex irreversible proofreading systems by mapping to the general scheme shown in Fig.~\ref{fig:genscheme}A. In particular, this includes translational proofreading systems with additional transitions both before and after the energy-consuming step.
	 Such schemes (see Fig.~\ref{fig:genscheme}B as an example) have been used to summarize experimental data on the kinetics of translational elongation in past work~\cite{rodnina2001ribosome,johansson2008kinetics,wohlgemuth2011evolutionary,zaher2010hyperaccurate}. The generalized scheme considered here uses an analogous coarse-graining procedure 
	to that described in recently published work by Igoshin, {\em et al}~\cite{igoshin2025coarse}, which trims states and non-energy-consuming loops using splitting probabilities and mean-first-passage times between remaining milestone states.

	In Fig.~\ref{fig:genscheme}A, the one-way arrows denote reactions where the reverse rates are negligible in their effect on splitting probabilities. The red arrows mark the only energy-consuming steps (eg: GTP hydrolysis), and the energy-consumption and release arrows denote transitions along distinct pathways. Release from the \change{high-energy} (purple) macrostate is assumed to return the system to the same (green) microstate, regardless of which tRNA is released.

	This system can be treated as a heterogeneous continuous-time random walk~\cite{grebenkov2018heterogeneous} with Markovian (albeit not constant-rate) transitions on a simplified coarse-grained network of states. We can define the splitting probabilities $p_c, p_w$ for transitioning out of the R into the RC$^*$ or RW$^*$ states respectively. Specifically, $p_c$ is the probability for a particle starting in the initial (green) state to first reach the compound state RC$^*$ before it ever reaches state RW$^*$. %For the Hopfield model in Fig.~\ref{fig:scheme}, these probabilities are given by $p_c = p_{bc}p_{hc}$ and $p_w = p_{bw}p_{hw}$.
	We can also define the probability $p_{pc}$ that a particle in the RC$^*$ macrostate will first transition to the elongated state, before a release occurs, and the probability $p_{uc} = 1-p_{pc}$ for the opposite case. 
	
	With these definitions, we can proceed using the same analysis as before. \change{Consider each individual `interaction' event with no intermediate returns to the initial state. The probability that an interaction passes through the RC$^*$ state} and results
in elongation is $p_c p_{pc}$. The probability that it passes through the RC$^*$ state but results in release is $p_c p_{uc}$.
	
	The error rate $f$ and the number of excess transitions over the energy-consuming pathway $\left<n\right>$, can then be  written analogously to Eq.~\ref{eq:avgn},~\ref{eq:fidelity}:
	
	\begin{subequations}
	\begin{align}
	f & = \frac{p_w p_{pw}}{p_c p_{pc}}, \label{eq:fgeneral}\\
	\left<n\right> & = \frac{p_c p_{uc} + p_w p_{uw}}{p_c p_{pc} + p_w p_{pw}}
	\end{align}
	\label{eq:fngeneral}
	\end{subequations}
	
	We note that the scheme in Fig.~\ref{fig:genscheme}A is equivalent to a substrate-selective Michaelis-Menten enzymatic reaction. For such reactions, the accuracy has previously been expressed as a ratio of the catalytic efficiencies $k_\text{cat}/K_\text{m}$ for the cognate versus noncognate substrates~\cite{johansson2012genetic,lovmar2006rate}. The error rate in Eq.~\ref{eq:fgeneral} is directly equivalent to such an expression.

	\subsection{Realistic model and parameters for ribosomal translation}
	For the specific scheme illustrated in  Fig.~\ref{fig:genscheme}B, employed in  Ref.~\cite{rodnina2001ribosome,wohlgemuth2011evolutionary}, the ``R" macrostate can be considered to include the initial binding and codon recognition transitions, as well as GTPase activation, with the energy-consuming exit from this macrostate corresponding to GTP hydrolysis. Since the GTPase activation is assumed to be effectively irreversible, the splitting probability out of the initial compound state is simply the probability that \change{hydrolysis} with the correct tRNA occurs before \change{hydrolysis} with the wrong tRNA (state 3C is reached before state 3W).
	This  probability can be computed through coarse-graining of the reaction scheme as follows.
	
	First, we find the probability $\hat{p}_{02}$ that a system starting at state 0 hits the 2C state before the 2W state. This can be done by considering each time the system leaves state $0$ as the start of an independent path. Each such path must either reach state $2C$ (with probabilistic weight $p_{01}^{(c)}p_{02}^{(c)}$)  or state $2W$ (with probabilistic weight $p_{01}^{(w)}p_{02}^{(w)}$) or else return to the beginning at state $0$. The resulting probability of hitting $2C$ first is then:
		\begin{equation}
	\begin{split}
	\hat{p}_{02}^{(c)} = \frac{p_{01}^{(c)}p_{12}^{(c)}}{p_{01}^{(c)}p_{12}^{(c)} + p_{01}^{(w)}p_{12}^{(w)}}, 
	\end{split}
	\end{equation}
	where $p_{ij}^{(c)}$ is the splitting probability from state $i$ to adjacent state $j$ with the correct tRNA. 
	Similarly, we find the probability $\hat{p}_{23}$ that a system starting at state 2 hits the 3C state before returning to the 0 state. Again we consider the weight of each path leaving state 2 without returning to it, to get:
	\begin{equation}
	\begin{split}	
	\hat{p}_{23}^{(c)} = \frac{p_{23}^{(c)}}{p_{23}^{(c)} + p_{21}^{(c)}p_{10}^{(c)}}.
	\end{split}
	\end{equation}
	Analogous probabilities are defined for the wrong tRNA. The desired splitting probability for leaving the macrostate entirely (through GTPase activation and hydrolysis) is then
	\begin{equation}
	\begin{split}	
	p_c = \frac{\hat{p}_{02}^{(c)}\hat{p}_{23}^{(c)}}{\hat{p}_{02}^{(c)}\hat{p}_{23}^{(c)} + \hat{p}_{02}^{(w)}\hat{p}_{23}^{(w)}}.
	\end{split}
	\end{equation}
	Note that this approach can be generalized to recursively compute splitting probabilities for any number of interchanging states arranged in a linear array. 
	
	\change{To obtain realistic parameters for ribosomal translation, we use the values reported in Ref.~\cite{wohlgemuth2011evolutionary} for a specific pair of cognate and near-cognate codons (UUU and CUC). The measured values are given in Fig.~\ref{fig:genscheme}B. Individual rates that were not measured (such as the EF-Tu shape change) follow a prior irreversible transition and do not affect the splitting probabilities of the system. In addition to the rates for a specific `wrong' tRNA, we need an estimate for the factor $g$, defining the concentration excess of wrong versus right substrates. A variety of different error rates have been measured for various near-cognate codon and anti-codon pairings~\cite{kramer2007frequency,daviter2006ribosome}, while non-near-cognate interactions are almost never observed~\cite{rodnina2001ribosome,joshi2019problem}. We estimate the relevant concentration factor by noting that  tRNA carrying the amino acid Lys has been shown to interact with $16$ cognate and near-cognate codons, with varying yet measurable frequencies~\cite{kramer2007frequency}. If each tRNA has some cross-reaction with $1/4$ of the possible codons, this would imply that each codon position has a measurable chance of incorporating $\sim 5$ of the twenty possible amino acids into the chain, through cognate and near-cognate tRNA interactions. 
		Of course, this estimate neglects the varying frequencies of distinct tRNAs {\em in vivo}, and the variety of translational kinetic rates for each. However, we take a rough average estimate of $g=4$ for the excess concentration of near-cognate tRNAs bearing the wrong amino acid that can interact with the ribosome sufficiently to proceed through the hydrolysis step and thereby contribute to the error and energetic cost.

	Using the numbers in Fig.~\ref{fig:genscheme}B~\cite{wohlgemuth2011evolutionary}, together with $g=4$, gives the estimated splitting probabilities $p_c=0.94, p_w = 0.06$ for passing the hydrolysis step with the right or wrong amino acid. The probabilities of elongation rather than release after hydrolysis are estimated as $p_{pc} = 0.98, p_{pw} = 0.04$, giving an overall error rate of $f \approx 2 \times 10^{-3}$ and a low cost in futile hydrolysis cycles of $\left<n\right> \approx 0.08$. We note that these numbers imply that $96^\%$ of wrong tRNAs are released after hydrolysis, while only $2^\%$ of correct tRNAs are released. Thus, this system is in a regime where much of the discrimination occurs on the initial binding step and most of the correct tRNAs that make it through hydrolysis go on to proceed to elongation, with very few futile cycles.

As shown in Fig.~\ref{fig:genscheme}C,D, this more realistic model of ribosome kinetics engenders a trade-off between energetic cost and accuracy that is analogous to the simple model described in the main text. When the post-hydrolysis rejection rate is low, the error rate is determined entirely by the initial steps leading to hydrolysis. We can define a discrimination factor for this passive part of the process as: 
\begin{equation}
e^{\tilde{\Delta}_1} = \log\left[p_{12}^c\hat{p}_{23}^c/\left(p_{12}^w\hat{p}_{23}^w\right)\right].
\end{equation}
  For the simple model in Fig.~\ref{fig:scheme}, the discrimination factor $\tilde{\Delta}_1$ approaches the binding energy difference $\Delta_1$ when hydrolysis is slow. For the realistic model in Fig.~\ref{fig:genscheme}B, where  hydrolysis  is relatively fast compared to unbinding, the parameter $\tilde{\Delta}_1 \approx 4.1$ serves an analogous role in setting the error rate in the absence of active proofreading  ($f \rightarrow g e^{-\tilde{\Delta}_1}$).
	
	For the discrimination factor associated with the proofreading step in this generalized model, we define
	\begin{equation}
	e^{\tilde{\Delta}_2} = \log \left[\left(k_{\text{rej,w}}/k_\text{pw}\right)/\left(k_\text{rej,c}/k_\text{pc}\right)\right].
	\end{equation}
	 This factor incorporates both the difference in release rate (as in the original simple model) and the difference in  elongation rate for the correct versus the wrong tRNA. For the measured parameters, it can be estimated as $\tilde{\Delta}_2\approx 7.5$. In the limit where rejection rates are much higher than elongation rates , the overall error approaches $f \rightarrow g e^{-\tilde{\Delta}_1 - \tilde{\Delta}_2}$. 
		
	The cost in terms of excess hydrolysis events is related to the error rate in much the same way as for the simple model. As plotted in Fig.~\ref{fig:genscheme}D, approaching the minimal possible error rate drives up the cost. There is a plateau  for intermediate error rates, corresponding to the regime where most wrong tRNAs are rejected while most correct ones move on to elongation. The plateau cost is equal to the passive error rate  ($ge^{-\tilde{\Delta}_1}$), and the plateau is wider for larger values of the proofreading discrimination factor $\tilde{\Delta}_2$. For the estimated rate constants used here, the system sits within a plateau region, with moderate error rates and relatively little waste in terms of futile hydrolysis events. This is consistent with past analyses, which indicated that ribosomal translation is optimized for speed and low energy cost rather than maximal accuracy~\cite{banerjee2017elucidating}.

	Given the analogous behavior of this more realistic model and the classic scheme in Fig.~\ref{fig:scheme}, we use the parameters $g = 4, \Delta_1 = 4, \Delta_2 = 8$ as the most relevant default values for analyzing the highly simplified model in the main text.}

\section{Appendix B: Fidelity and energetic cost for reversible model}	

For the model with reversible transitions (Fig.~\ref{fig:energylandscape}) we can define the splitting probabilities for correct and wrong tRNAs \change{in terms of either the rate constants or the transition state energies, as follows}:
\begin{subequations}
	\begin{align}	
		p_{bc} & = \frac{k_b}{(k_b+k_{ur})(1+g)} = \frac{e^{-E_b^\ddagger}}{(e^{-E_b^\ddagger} + e^{-E_2^\ddagger})(1+g)}, 
		& \; p_{bw} & = \frac{gk_b}{(k_b+k_{ur})(1+g)} = \frac{ge^{-E_b^\ddagger}}{(e^{-E_b^\ddagger} + e^{-E_2^\ddagger})(1+g)},\\
		p_{urc} & = \frac{k_{ur}}{(k_b+k_{ur})(1+g)} = \frac{e^{-E_2^\ddagger}}{(e^{-E_b^\ddagger} + e^{-E_2^\ddagger})(1+g)}, 
		& \; p_{urw}  & = \frac{gk_{ur}}{(k_b+k_{ur})(1+g)} = \frac{ge^{-E_2^\ddagger}}{(e^{-E_b^\ddagger} + e^{-E_2^\ddagger})(1+g)},\\
		p_{hc} & = \frac{k_h}{k_h  + k_{u1}} = \frac{e^{-E_h^\ddagger + \epsilon_\text{drive}}}{(e^{-E_h^\ddagger+\epsilon_\text{drive}} + e^{-E_b^\ddagger})}, 
		\;  & p_{hw} & = \frac{k_h}{k_h  + k_{u1w}} = \frac{e^{-E_h^\ddagger + \epsilon_\text{drive}}}{(e^{-E_h^\ddagger+\epsilon_\text{drive}} + e^{-E_b^\ddagger+\Delta_1})},\\
		p_{rc} & = \frac{k_r}{k_r + k_{u2} + k_{p}} = \frac{e^{-E_h^\ddagger}}{(e^{-E_h^\ddagger} + e^{-E_2^\ddagger} + e^{-E_p^\ddagger})}, 
		&  \; p_{rw}  & = \frac{k_{rw}}{k_{rw} + k_{u2w} + k_{p}} =  \frac{e^{-E_h^\ddagger-\Delta_1}}{(e^{-E_h^\ddagger-\Delta_1} + e^{-E_2^\ddagger} + e^{-E_p^\ddagger-\Delta_2})}, \\
		p_{pc} & = \frac{k_p}{k_r + k_{u2} + k_{p}} =  \frac{e^{-E_p^\ddagger}}{(e^{-E_h^\ddagger} + e^{-E_2^\ddagger} + e^{-E_p^\ddagger})}, 
		& \; p_{pw}  & = \frac{k_p}{k_{rw} + k_{u2w} + k_{p}} = \frac{e^{-E_p^\ddagger-\Delta_2}}{(e^{-E_h^\ddagger-\Delta_1} + e^{-E_2^\ddagger} + e^{-E_p^\ddagger-\Delta_2})},
	\end{align}
\end{subequations}
where $k_h = k_h^0 + \alpha = k_{h}^0 e^{\epsilon_\text{drive}}$ is the total hydrolysis rate.

We consider individual interactions of a tRNA with a ribosome, each of which involves leaving the empty $R$ state of the ribosome and eventually returning to it (possibly with a longer peptide chain), without any intermediate visits to that state. Each interaction can be resolved through either unbinding, release from the \change{high-energy intermediate} state, or elongation. 	

For a system that reaches the \change{high-energy} $RC^*$ state, the probabilistic weight of all paths with exactly $i$ \change{hydrolysis} transitions since the start of the interaction is:
\begin{equation}
	\begin{split}	
		w_{i,c} & =  p_{urc}(p_{rc}p_{hc})^i + 
		p_{bc}p_{hc}(p_{rc}p_{hc})^{i-1}, \\	
	\end{split}
\end{equation}
with an analogous expression $w_{i,w}$ for a system in the $RW^*$ state. The probability that the interaction resolves in elongation with the correct or the wrong tRNA ($p_{el,c}, p_{el,w}$) is then
\begin{equation}
	\begin{split}	
		p_\text{el,c} & = p_{urc}p_{pc} + \sum_{i=1}^\infty w_{ic}p_{pc} = \frac{(p_{urc}+p_{bc}p_{hc})p_{pc}}{1-p_{rc}p_{hc}} \\
		p_\text{el,w} & = \frac{(p_{urw}+p_{bw}p_{hw})p_{pw}}{1-p_{rw}p_{hw}}.
	\end{split}
\end{equation}
The error rate is given by $f = p_\text{el,w}/p_\text{el,c}$, yielding Eq.~\ref{eq:f_reversible}.

The average number of \change{hydrolysis} steps for an interaction involving the correct tRNA ($N_c$) is found by summing over the corresponding $w_{i,c}$, multiplied by the probability that the \change{high-energy} state resolves with no further \change{hydrolysis} transitions: $p_{fc} = 1 - p_{rc} p_{hc}$. The result for both correct and wrong interactions is:
\begin{equation}
	\begin{split}
		N_c & = \sum_{i=0}^\infty i w_{i,c} p_{fc} = \frac{(p_{bc} + p_{urc}p_{rc})p_{hc}}{1-p_{rc}p_{hc}} \\
		N_w & = \frac{(p_{bw} + p_{urw}p_{rw})p_{hw}}{1-p_{rw}p_{hw}}
	\end{split}
\end{equation}
Finally, we can find $\left<N\right>$: the average number of driven activation transitions per interaction event, conditional on that event resolving to elongation. Here we multiply by the fraction of activation transitions that proceed along the driven arrow rather than the basal activation process: $\alpha/(k_h^0 + \alpha) = 1-e^{-\epsilon_\text{drive}}$, to yield:
\begin{equation}
	\begin{split}	
		\left<N\right> = \left[\frac{N_c + N_w}{p_{el,c} + p_{el,w}}\right](1-e^{-\epsilon_\text{drive}}),
	\end{split}
\end{equation}
which gives Eq.~\ref{eq:N_reversible} in the main text.

\section{Appendix C: Catalytic control from exploratory dynamics with resetting}

Below we describe two distinct approaches to computing the steady-state length distribution of a simple model consisting of reversible motion along a linear set of states, with a constant (catalytically controlled) rate of resetting to the origin. The first approach uses weighted path counting, analogously to the proofreading models in the main text, to compute the length at which resetting occurs. This approach is simple enough to compute manually, but is limited to irreversible resetting transitions. The second approach maps the system to an electric circuit, which can be analyzed in its entirety using well-established matrix methods.

\subsection{Microtubule length control by irreversible resetting, via path-counting}
Here we demonstrate how the approach of adding up probabilistically weighted paths can be extended to larger reaction systems with multiple intermediate states. The system considered here is a simple model of dynamic instability (Fig.~\ref{fig:catalytic}A), in the limit where the catastrophe transitions are effectively irreversible. For a microtubule that initially starts at length $1$, We seek to compute the probability $P^\text{cat}_L$ that catastrophe occurs from the $L$ state. For simplicity, we assume that the forward, reverse, and catastrophe rates are constant for all states.
		
We begin by defining the splitting probabilities at each state. For state $i>1$, the probabilities of stepping forward and backward are, respectively, $p_f = (k_f+\alpha)/(k_f+\alpha+k_r+k_c), p_r = k_r/(k_f+\alpha+k_r+k_c)$. From each such state there is also a catastrophe probability $p_c = k_c/(k_f+\alpha+k_r+k_c)$. For state $i=1$, the probability of stepping forward is simply 1. In the case where $p_r \rightarrow 0$, the system steps forward until catastrophe is reached. The probability this happens in state $i$ is then the product of forward-stepping probabilities for the first $i-1$ steps times the catastrophe probability:  $P^\text{cat}_i = p_f^{i-2}(1-p_f)$. This probability is normalized over $i>1$. 
		
For the case with substantial reversals, the paths can include any number of back and forth steps, so long as they never go below 0 and end in the $i^\text{th}$ state.	These paths can be conveniently enumerated using stone-fence diagrams and continued fractions, as previously employed for computing statistics of semiflexible polymers~\cite{yamakawa1973statistical, spakowitz2005end}. Specifically, we define $w_i^+$ as the total weight of all paths that start at state $i$ and never go below it. This quantity can be expanded as
\begin{equation}
	\begin{split}
		w_i^+ & = 1 + p_f w_{i+1}^+ p_r + (p_f w_{i+1}^+ p_r)^2 + \ldots
			= \frac{1}{1 - p_f p_r w_{i+1}^+}, \quad i>1, \\
		w_1^+ & = 	\frac{1}{1 - p_r w_{i+1}^+}.			
	\end{split}
\end{equation}
Here, the first term is a path of length $0$, the second includes all paths that step up from the $i$th level only once (but can meander arbitrarily at $i+1$ and above), the second term corresponds to paths that step up above the $i$th level twice, and so on. Because we allow the states to go infinitely high, we must have $w_i^+ = w^+$, a constant for all states $i>1$. We can then get the closed-form expression:
\begin{equation}
	\begin{split}
		w^+ = (1 - \sqrt{1-4p_fp_r})/(2p_fp_r).
	\end{split}
\end{equation}		
		
\begin{figure}
	\centerline{\includegraphics[width=0.47\textwidth]{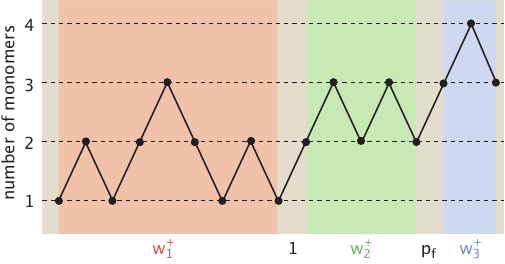}}
	\caption{Stone-fence diagram illustrating example path starting in state $1$ and ending in state $i=4$. Levels correspond to the sequential state $i$ (eg: length of a microtubule in monomers). The path is decomposed into components that end at the last visit to each level, and the probabilistic weights of each component are given beneath.
	}
	\label{fig:stonefence}
\end{figure}		
		
Any path starting at $1$ and ending at state $i$ (with no intervening catastrophes) can be decomposed into the following sequential components: the part of the path up to its last visit at $1$, then a step up to $2$, the next part of the path up until its last visit at $2$, then a step up to $3$ and so on until you reach the last section of the path that starts and ends at $i$ and never goes beneath it. This decomposition for an example path is illustrated in Fig.~\ref{fig:stonefence}. The resulting total weight is then multiplied by $p_c$ to compute the probability that catastrophe happens at state $i$:
\begin{equation}
	\begin{split}	
		P^\text{cat}_i & = (w_1^+\cdot 1 \cdot w_2^+ \cdot p_f \cdot w_3^+ \cdot p_f \ldots \cdot w_i^+) p_c \\
		& = p_c w_1^+w^+ (w^+ p_f)^{i-2} 
		 = (1 - w^+p_f)(w^+ p_f)^{i-2},
	\end{split}
\end{equation} 
where the last expression accounts for the normalization of the probabilities added up over all states from 2 onwards.

This geometric series skews towards longer lengths when the forward stepping probability $p_f$ becomes high. This probability represents a balance between the forward stepping rate versus reversal or catastrophe. At equilibrium, the catastrophe can only be effectively irreversible if the energies associated with longer-length states are much higher than shorter ones, implying $k_f / k_r \ll 1$. If there is little driving in the system ($\alpha \ll k_r$) then this means $p_f \rightarrow 0$ and the distribution of lengths becomes peaked at $1$, regardless of the rate of catastrophe. On the other hand, if the system is strongly driven, then $w^+\rightarrow 1$ and we recover the limit with unidirectional stepping, discussed above. The distribution is then determined by $p_f \approx \alpha/(\alpha+k_c)$, with the average length at catastrophe given by $\left<L_\text{cat}\right> \rightarrow (2-p_f)/(1-p_f) \rightarrow \alpha/k_\text{cat}$ for large $\alpha$. Thus, active driving allows the steady-state microtubule length to be linearly sensitive to the level of catalyst present, ensuring catalytic control.
		
Finally, we note that because the catastrophe rate is the same from each state, the distribution of length upon catastrophe is directly proportional to the steady-state length distribution: $P_i = \gamma P_i^\text{cat} $, where $\gamma = 1-P_1$ is an appropriate normalization constant.

\subsection{Microtubule catalytic control via the circuit mapping}
In terms of the $n$th mesh current shown in Fig. ~\ref{fig:catalytic}B, the voltage equation taken along the path of the $n$th battery is:
\begin{equation}
P_{n+1}e^{\beta G_{n+1}}-P_{n}e^{\beta G_n}= {{\alpha}\over{k_f}}P_n e^{\beta G_n} - R_n I_n.
\end{equation}
where $R_n = e^{\beta G_n}/k_f$ and $G_n=nG$. Note that $e^{\beta G} = k_b/k_f$, where $k_f$ and $k_b$ are the equilibrium forward and backward rates, respectively. Using these definitions, we can solve for the probability of the $(n+1)$th state in terms of the previous state probability and current:
\begin{equation}
P_{n+1}=\Big(1+{{\alpha}\over{k_f}} \Big) e^{-\beta G}P_n-I_n {{e^{-\beta G}}\over{k_f}}
\end{equation}

Taking the potential difference from state $n+1$ and state 1 along the catastrophe path:
\begin{equation}
P_{1}e^{\beta G_{1}}-P_{n+1}e^{\beta G_n+1}= -R_{\mathrm{cat}, n} (I_{n}-I_{n+1}),
\end{equation}
Where $R_{\mathrm{cat}, n} = e^{\beta G_{n+1}}/k_{\mathrm{cat}} = R_{n+1} ({{k_f}\over{k_{\mathrm{cat}}}})$.
Therefore, the $(n+1)$th current is:
\begin{equation}
I_{n+1}= I_n - k_{\mathrm{cat}}P_{n+1}+ k_{\mathrm{cat}}P_{1}e^{-\beta G_{n}}
\end{equation}

In vector notation, the recursive probability and current equations become:
\begin{equation}
\begin{bmatrix}
    1       & 0 \\
    k_{\mathrm{cat}}       & 1
\end{bmatrix}
\begin{bmatrix}
    P_{n+1} \\
    I_{n+1} 
\end{bmatrix}
=
\begin{bmatrix}
    \Big(1+{{\alpha}\over{k_f}} \Big) e^{-\beta G}       & -{{e^{-\beta G}}\over{k_f}} \\
    0      & 1
\end{bmatrix}
\begin{bmatrix}
    P_{n} \\
    I_{n} 
\end{bmatrix}
+
\begin{bmatrix}
    0 \\
    {{k_{\mathrm{cat}}P_{1}}\over{e^{\beta G_{n}}}}
\end{bmatrix}
\end{equation}
Multiplying both sides by the inverse of the right-hand-side matrix, the recursion relation is:
\begin{equation}
\begin{bmatrix}
    P_{n+1} \\
    I_{n+1} 
\end{bmatrix}
=
M
\begin{bmatrix}
    P_{n} \\
    I_{n} 
\end{bmatrix}
+
\begin{bmatrix}
    0 \\
    {{k_{\mathrm{cat}}P_{1}}\over{e^{\beta G_{n}}}}
\end{bmatrix}
\end{equation}
where the transition matrix $M$ is given by:
\begin{equation}
M=
\begin{bmatrix}
    \Big(1+{{\alpha}\over{k_f}} \Big) e^{-\beta G}       & -{{e^{-\beta G}}\over{k_f}} \\
    -k_{\mathrm{cat}}\Big(1+{{\alpha}\over{k_f}} \Big) e^{-\beta G}      & k_{\mathrm{cat}}{{e^{-\beta G}}\over{k_f}}+1
\end{bmatrix}
\end{equation}

The probability and current of state $n$ in terms of those of state 1 is thus:
\begin{equation}
\begin{bmatrix}
    P_{n+1} \\
    I_{n+1} 
\end{bmatrix}
=
M^n
\begin{bmatrix}
    P_{1} \\
    I_{1} 
\end{bmatrix}
+\sum_{k=0}^{n-1}(e^{\beta G}M)^k
\begin{bmatrix}
    0 \\
    {{k_{\mathrm{cat}}P_{1}}{e^{-\beta G n}}}
\end{bmatrix}
\end{equation}
Diagonalizing $M$:
\begin{equation}
M=V
\begin{bmatrix}
    \lambda_-       & 0 \\
    0      & \lambda_+
\end{bmatrix}
V^{-1}
\end{equation}
Where the columns of $V$ are the eigenvectors of $M$ and $\lambda_-$ and $\lambda_+$ are the eigenvalues of $M$:
\begin{equation}
\lambda_{\pm} = {{e^{-\beta G}}\over{2k_f}} \Bigg( \alpha+ k_f(1+e^{\beta G})+k_{\mathrm{cat}} \pm \sqrt{(\alpha+k_f-k_fe^{\beta G})^2+k_{\mathrm{cat}}(2\alpha+2(1+e^{\beta G})k_f+k_{\mathrm{cat}})}\Bigg)
\end{equation}
Which simplifies to the value given in the text:
\begin{equation}
D=-\ln{\lambda_{-}} = -\ln{[1- {{\sqrt{(\alpha+k_{\mathrm{cat}}+k_f-k_r)^2+4k_{\mathrm{cat}}k_r}-(\alpha+k_{\mathrm{cat}}+k_f-kr)}\over{2k_r}}]} 
\end{equation}

Note that $\lambda_- \leq 1$ whereas $\lambda_+ \geq 1$.

The transfer matrix equation is then
\begin{equation}
\begin{bmatrix}
    P_{n+1} \\
    I_{n+1} 
\end{bmatrix}
=
V
\begin{bmatrix}
    \lambda_-^n       & 0 \\
    0      & \lambda_+^n
\end{bmatrix}
V^{-1}
\begin{bmatrix}
    P_{1} \\
    I_{1} 
\end{bmatrix}
+\sum_{k=0}^{n-1}e^{\beta k G}
V
\begin{bmatrix}
    \lambda_-^k       & 0 \\
    0      & \lambda_+^k
\end{bmatrix}
V^{-1}
\begin{bmatrix}
    0 \\
    {{k_{\mathrm{cat}}P_{1}}{e^{-\beta G n}}}
\end{bmatrix}
\end{equation}
Expanding this expression and taking the geometric sum yields $P_{n}$:
\begin{equation}
P_{n}= P_1e^{-\beta G(n-1)} {{k_{\mathrm{cat}}}\over{k_{\mathrm{cat}}-\alpha (e^{\beta G}-1)}} + A_1 \lambda_-^{n-1} + A_2 \lambda_+ ^{n-1},
\end{equation}
where the $A_1$ and $A_2$ are explicit functions of the elementary parameters. For nonzero $k_{\mathrm{cat}}$ the probability monotonically decreases for larger $n$, thus the coefficient $A_2$ must be zero. Solving this boundary condition for $I_1$ and substituting into the expression for $A_1$, we obtain the length distribution (Eq. 14 in the main text):

\begin{equation}
P_{n}= {{k_{\mathrm{cat}}}\over{k_{\mathrm{cat}}-\alpha (e^{\beta G}-1)}}P_1e^{-\beta G(n-1)}+{{\alpha(e^{\beta G}-1)}\over{\alpha (e^{\beta G}-1)-k_{\mathrm{cat}}}}P_1 e^{-D(n-1)},
\end{equation}
where $D=-\ln{\lambda_-}$.

\change{For microtubule assembly, the net growth rate at physiological $\alpha$ is  $\Delta x (k_f+\alpha - k_r) = 10.4$ microns per second, where the change in length per addition of tubulin dimer $\Delta x = 8/13$ nanometers because each tubulin dimer is 8 nanometers long and 13 dimers complete a single turn of the microtubule helix. Microtubule growth and decay rates are assumed to be equal for interphase and mitosis, which are the same to within experimental noise. The catastrophe rate $k_{\mathrm{cat}}$ equals 0.01 per second and 0.1 per second in interphase and mitosis, respectively. The mean filament lengths are 23 (11) and 6(3) microns in interphase and mitosis, respectively. All of these parameters and measurements were obtained from Ref. \cite{belmont1990real}, which reported values for Xenopus egg extract with different amounts of Rhodamine tubulin added; for consistency within our model, we used parameters corresponding to 1mg/ml Rhodamine tubulin. $k_r$ was obtained using the values for tubulin bound to the non-hydrolyzable proxy substrate GMPCPP reported in \cite{microtubule_howard}. }

From the expression for $D$, we can see that the mean length and the sensitivity of the mean length to $k_{\mathrm{cat}}$ is maximal in the limit that $4k_{\mathrm{cat}}k_r / (\alpha + k_{\mathrm{cat}}+k_f-k_r)^2 << 1$(visualized in Fig. ~\ref{fig:catalytic}C). Expanding the expression for $D$ to first order in this ratio, we obtain:
\begin{equation}
D \approx -\ln{\Bigg[1-{{k_{\mathrm{cat}}}\over{|k_{\mathrm{cat}}+\alpha+k_f-k_r|}}\Bigg]} 
\end{equation}
In this limit, the mean length retains linear sensitivity to $k_{\mathrm{cat}}$ (that is, the linear approximation to the logarithm is valid) if $\alpha > k_r-k_f + \Delta$, where the minimum buffer $\Delta$ is set by the value of $k_{\mathrm{cat}}$ because ${{k_{\mathrm{cat}}}\over{k_{\mathrm{cat}}+\Delta}}$ must be much less than 1. Therefore, as stated in the main text, the transition from weak (logarithmic) to strong (linear) catalytic regulation occurs when $\alpha > k_r - k_f$, with the sharpness being inversely proportional to $k_{\mathrm{cat}}$.
        
\end{document}